\documentstyle[12pt]{article}

\renewcommand{\theequation}{\arabic{section}.\arabic{equation}}

\newcommand{\be}{\begin{equation}}
\newcommand{\ee}{\end{equation}}
\newcommand{\bea}{\begin{eqnarray}}
\newcommand{\eea}{\end{eqnarray}}

\newcommand{\p}[1]{(\ref{#1})}% WAKimoto

\def\lefthook{{\vrule height5pt width0.4pt depth0pt}}
\def\righthook{{\vrule height5pt width0.4pt depth0pt}}
\def\leftrighthookfill{$\mathsurround=0pt \mathord\lefthook
     \hrulefill\mathord\righthook$}
\def\underhook#1{\vtop{\ialign{##\crcr$\hfil\displaystyle{#1}\hfil$\crcr
      \noalign{\kern-1pt\nointerlineskip\vskip2pt}
      \leftrighthookfill\crcr}}}

\newcommand{\dif}{\partial}

\font\ninerm=cmr9

\begin{document}
\thispagestyle{empty}
\setcounter{page}{0}
\renewcommand{\thefootnote}{\fnsymbol{footnote}}
\fnsymbol{footnote}

%\phantom{Nothing}
\rightline{LBNL-39562}
\rightline{UCB-PTH-96/49}
\rightline{BONN-TH-96/16}
\rightline{hep-th/9611083}
\rightline{November 1996}

\vspace{.25in}

\begin{center} \Large \bf Wakimoto  realizations of  current algebras:\\

an explicit construction\\

\end{center}

\vspace{.2in}

\begin{center}
Jan de Boer${}^\dagger$ 
and L\'aszl\'o Feh\'er${}^*$\\

\vspace{0.3in}

{\em 
${}^\dagger$Department of Physics, University of California at Berkeley\\
366 LeConte Hall, Berkeley, CA 94720-7300, U.S.A.\\
and\\
Theoretical Physics Group, Mail Stop 50A-5101\\
Ernest Lawrence Berkeley Laboratory, Berkeley, CA 94720, U.S.A.\\
e-mail: deboer@theor3.lbl.gov\\

\vspace{0.2in}

${}^*$Physikalisches Institut, Universit\"at Bonn\\
Nussallee 12, D-53115 Bonn, Germany\\
e-mail: feher@avzw01.physik.uni-bonn.de
}

\end{center}

\vspace{.3in}

{\parindent=25pt
\noindent

\begin{center} {\bf Abstract} \end{center}

\small

A generalized Wakimoto realization  of $\widehat{\cal G}_K$  can be  
associated with each parabolic subalgebra 
${\cal P}=({\cal G}_0 +{\cal G}_+)$ 
of a simple Lie algebra ${\cal G}$ according  to an earlier proposal by  
Feigin and Frenkel.  
In this paper the  proposal is made explicit  by developing the construction 
of  Wakimoto realizations  from a simple but unconventional viewpoint. 
An explicit formula is  derived for the Wakimoto  current first at the 
Poisson bracket level by Hamiltonian symmetry reduction of the WZNW model.
The  quantization is then performed  by normal ordering the classical
formula and determining the required quantum correction for it to
generate $\widehat{\cal G}_K$ by means of commutators.
The affine-Sugawara stress-energy tensor is verified to have the expected
quadratic form in the constituents, which are symplectic bosons 
belonging to ${\cal G}_+$ and a current belonging to ${\cal G}_0$.
The quantization requires a  choice of  special polynomial coordinates 
on the big cell of the flag manifold $P\backslash G$. 
The effect of this choice is investigated in detail by constructing
quantum coordinate transformations.
Finally, the explicit form of the screening charges for each generalized
Wakimoto realization is determined, and some applications are briefly 
discussed.
}

\normalsize

\newpage

\renewcommand{\thefootnote}{\arabic{footnote}}
\setcounter{footnote}{0}

\section{Introduction}

%%%%%%%% SOME USEFUL ABBREVIATIONS %%%%%%
\def\al{\alpha}
\def\bt{\beta}
\def\d{\delta}
\def\gm{\gamma}
\newcommand{\g}{\gamma}
\def\Om{\Omega}
\def\ttt{\tilde{T}}
\def\nonu{\nonumber \\ {} }
\def\p{\phi}
\def\v{\varphi}
\def\la{\langle}
\def\ra{\rangle}
\def\pa{\partial}
\def\s{\vskip 5mm}
\def\i{\int d^2x\,}
\def\hb{\hfill\break}
\def\t{{\rm Tr}\,}
\def\tr{{\rm Tr}\,}
\def\dg{\delta g\,g^{-1}}
\def\gd{g^{-1}\delta g}
\def\fg{{{\d F}\over{\d g}}}
\def\fj{{{\d F}\over{\d J}}}
\def\hg{{{\d H}\over{\d g}}}
\def\hj{{{\d H}\over{\d J}}}
\def\si{\sigma}
\def\bsi{\bar\sigma}
\def\D{{\cal D}}
\def\G{{\cal G}}
\def\W{{\cal W}}
\def\S{{\cal S}}
\def\V{{\cal V}}

Realizations of various symmetry algebras in terms of simpler algebras
are often useful in representation theory and in physical applications.
It is enough to mention, for example, the oscillator realizations
of simple Lie algebras, the Feigin-Fuks free boson realization 
of the Virasoro algebra, or 
the Miura transformations pertinent to the
theory of integrable hierarchies and ${\cal W}$-algebras.
This paper deals with another instance of this general theme,
the so-called  Wakimoto  realizations of the non-twisted affine
Lie algebras.

 Wakimoto \cite{Wa}  found the following  formula 
for the generating currents of $\widehat{sl}(2)_K$:
\be
I_- =-p,
\quad
I_0 =j_0 -2\left(pq\right),
\quad
I_+=-\left(j_0q\right)+K\dif q +\left(p(qq)\right)
\label{1}\ee
where   $j_0=-i\sqrt{2(K+2)}\dif \phi$ and the constituents 
are free fields subject to  the singular operator product 
expansions\footnote{We adopt  the
conventions of \cite{BBSS} for OPEs and for normal ordering.}
(OPEs)
$$%\be 
p\underhook{(z) q}(w) = \frac{-1 }{z-w},
\qquad
\dif\phi \underhook{(z) \dif \phi}(w) = \frac{-1}{(z-w)^2}.
$$%\label{2}\ee
The affine-Sugawara  stress-energy tensor is quadratic in the free fields
$$%\be
\frac{1}{2(K+2)}\left({\tr}I^2\right)=
-\frac{1}{2}
\left(\dif \phi \dif \phi\right) -\frac{i}{\sqrt{2(K+2)}}\dif^2\phi
-\left(p\dif q\right).
$$%\label{3}\ee
This free field realization raised considerable 
interest because of its usefulness in computing correlation
functions of conformal field theories \cite{FF,BF,BMP,ATY,Ku}
 and in analyzing  the 
quantum  Hamiltonian
reduction of $\widehat{sl}(2)_K$ to the Virasoro algebra \cite{BO}.

A detailed study of `Wakimoto realizations'
of arbitrary non-twisted affine Lie algebras 
was undertaken in \cite{FF,BF,BMP,GMOMS,KS,IKa,IKo,ATY,Ku}.
These  investigations  made use of 
the observation that at the level of zero modes (\ref{1}) becomes
the well-known differential operator realization of $sl(2)$ given by
%\be
$I_-\rightarrow {\dif \over \dif q}$,
%\qquad
$I_0\rightarrow -\mu_0 +2 q {\dif \over \dif q}$,
%\quad
$I_+\rightarrow \mu_0 q - q^2 {\dif \over \dif q}$,
%\qquad 
for some  $\mu_0\in {\bf C}$.
%\label{4}\ee
Any simple Lie algebra, $\G$, admits analogous 
realizations in terms of first order order  differential operators,  
\be
X\mapsto  \hat X=-F^X_\alpha(q) 
{\partial \over \partial q_\alpha}+ H^X_{\mu_0}(q),
\qquad \forall\, X\in \G,
\label{5}\ee
where $F_\alpha^X$, $H_{\mu_0}^X$ are polynomials 
(see e.g.~\cite{Ko,A}).
In the context of Wakimoto realizations mainly the so called 
`principal  case' was considered. 
In this case  $\alpha$ runs over the positive roots and $\mu_0$ is a
character of the Borel subalgebra of $\G$, which  may be identified 
with an element of the Cartan subalgebra ${\cal H}\subset \G$.
The main idea  \cite{FF,BMP,IKo}  for generalizing (\ref{1}) 
was  to  regard the differential operator realization of $\G$
as the zero mode part of the sought after realization of $\widehat{\G}_K$,
which should be obtained by `affinization' where
one replaces $q_\alpha$ and $p^\alpha$ by conjugate quantum fields and
$\mu_0$ by a current that (in the principal case) belongs to ${\cal H}$. 
Of course, one also needs to add derivative terms  and 
find the correct normal ordering, which is rather nontrivial.

Feigin and Frenkel  demonstrated 
 by indirect, homological 
techniques  that 
the affinization can be performed for any  simple Lie algebra $\G$ 
 \cite{FF} (see also \cite{Ku}).
An explicit  formula for the currents 
corresponding to the Chevalley generators of $sl(n)$ was given in \cite{FF} 
too, but  the method does  not lead to  explicit formulas 
for all currents of an  arbitrary $\G$ (or  $\G=sl(n)$).
Explicit formulas were later obtained in \cite{IKo}  for any $\G$, 
but again only for the Chevalley generators, and without complete proofs.
The quadraticity of the affine-Sugawara stress-energy tensor 
in the free fields is not quite transparent in this approach.

Elaborating the preceding  announcement \cite{dBF}, 
in  this paper we present an explicit  construction of the Wakimoto
realization.
The essence of our method is that we first derive a classical, 
 Poisson bracket
version of the Wakimoto realization by Hamiltonian reduction of the
WZNW model,  and then quantize the classical Wakimoto 
realization  by  normal ordering.
This procedure makes it possible to obtain  an  explicit 
formula for all currents of $\G$ 
in terms of finite  group theoretic data.

We shall achieve the above result  in the general  case, 
which was previously  investigated  by Feigin and Frenkel 
in their different approach (see the second article in \cite{FF}).
The general case is characterized  by  the choice of a parabolic
subalgebra,  ${\cal P}=(\G_0 +\G_+)\subset \G$.
Here $\G= \G_- + \G_0 + \G_+ $ denotes a  triangular decomposition 
induced by a  ${\bf Z}$-gradation of $\G$.
One also uses  a connected complex Lie group $G$ 
corresponding to $\G$,  its connected subgroups $G_{0}, G_\pm \subset G$
with Lie algebras $\G_{0}, \G_\pm\subset \G$, and 
the  parabolic subgroup $P=G_0 G_+$.
The principal case is that of the principal gradation of $\G$,  
for which $\G_0={\cal H}$ and  ${\cal P}$ is the Borel subalgebra.

Our starting point will be the  Hamiltonian interpretation 
of the differential operator realization of $\G$ mentioned in (\ref{5}). 
The base space of these differential operators 
is the big cell of the flag manifold $P\backslash G$,
and $\mu_0$ is taken to be a character of ${\cal P}$ if one requires 
the operators  to act on scalar, as opposed to vector, valued 
holomorphic functions. 
Identifying the big cell with the nilpotent subgroup $G_- \subset G$,
we  replace the differential operator realization with the 
equivalent Poisson bracket realization given on the 
holomorphic cotangent bundle 
$T^* G_-$ by 
$$
X\mapsto I^X(q,p)\equiv F^X_\alpha(q) p^\alpha + H^X_{\mu_0}(q)
$$
in terms of some  canonical coordinates $q_\alpha$, $p^\alpha$ on 
$T^* G_-$.
This formula can be  elegantly written as $I^X = \tr(XI)$ with 
$$
I= g_-^{-1} (-j - \mu_0) g_-,  
$$
where $g_-\in G_-$,  $j: T^* G_- \rightarrow \G_-^*=\G_+$
is the momentum map for the action of $G_-$ on $T^* G_-$ that comes from 
left translations, and 
$\mu_0\in  \G_0\simeq \G_0^*\subset {\cal P}^* $ defines
 a character, i.e., an element of $[{\cal P}, {\cal P}]^\perp$. 
Using  dual bases $V^\beta\in \G_+$ and  $V_\alpha\in \G_-$,
we have\footnote{Summation on coinciding indices is understood 
throughout the text.}
\be
j(q,p)={\cal F}_{\alpha\beta}(q)  p^\beta V^\alpha
\label{9}\ee
with  $-{\cal F}_{\alpha\beta}(q)  {\pa \over \pa q_\beta}$ 
being  the vector field on $G_-$ that is tangent to the curve 
$g_- \mapsto e^{- t V_\alpha} g_-$. 
As explained in \cite{dBF}, the above formula of  $I$, 
 which is equivalent to a formula
of Kostant \cite{Ko}  for the differential operators on $G_-$,
naturally follows from a Hamiltonian symmetry reduction on 
the holomorphic cotangent bundle $T^*G$.
The reduction uses  the action of the symmetry group $P$ on $T^* G$
that comes from left translations, and therefore the action of 
$G$ defined by right  translations survives to give an action of $G$ 
on the reduced space $T^* (P\backslash G)$, 
whose generator (momentum map) on the big cell 
$T^* G_- \subset T^* (P\backslash G)$ is just $I$.
The value of the 
momentum map that defines the reduction is given by 
$-\mu_0 \in \G_0=\G_0^* \subset {\cal P}^*$.
The reduction can be considered in the same way if 
$\mu_0\in \G_0$ is not necessarily a character but arbitrary, 
and then the formula of $I$ becomes
\be
I(q,p,j_0)= g_-^{-1} ( -j + j_0) g_- 
\label{10}\ee
where  $j_0$  is a variable  on the  co-adjoint 
orbit ${\cal O}$ of $G_0$ through $- \mu_0$. 
Remember that a character is a one-point co-adjoint orbit, 
and notice that a co-adjoint orbit of $G_0$ can  be regarded 
as a special co-adjoint orbit of $P$.
Formula (\ref{10}) is  valid on 
the big cell $T^* G_- \times {\cal O}$ in  the reduced phase 
space.\footnote{Globally, the reduced phase space is a fibre bundle 
over $T^*(P\backslash G)$ whose  fibre  is the co-adjoint orbit
${\cal O}$  of $P$, as follows from general results 
on reductions of cotangent bundles \cite{GS,Ro}.}

The Hamiltonian reduction on  $T^*G$ that is behind the Poisson
bracket realization of $\G$ given by (\ref{10}) can be 
naturally generalized to a reduction of the WZNW model \cite{Wi}  based 
on $G$.
As we shall see, this leads to the classical Wakimoto realization
of the affine Lie algebra $\widehat{\G}_K$ generated by the $\G$-valued
current
\be
I(q,p,j_0) = g_-^{-1} ( -j + j_0) g_-  + K g_-^{-1} g_-'
\label{11}\ee
where now all constituents have been promoted to be fields 
 on $S^1$.  
In particular,  $q_\alpha$, $p^\alpha$ that define $j$ 
by (\ref{9}) are now conjugate classical fields
and $j_0$ is a $\G_0$-valued current (belonging to a
co-adjoint orbit of the centrally extended loop group of $G_0$).

After presenting the derivation of formula (\ref{11}) 
for the classical Wakimoto current in Section 2, 
we shall quantize this formula explicitly.
For this we shall replace the constituent free fields 
$q_\alpha$, $p^\alpha$ and $j_0$ by corresponding quantum fields, 
 normal order the formula and determine an additional quantum
correction so that $I$ will indeed generate the current
algebra. We then  verify the expected quadratic form of the  
Sugawara expression. 
The result is given by  Theorem 3 in Section 3.
All previously known explicit formulas for the Wakimoto realization
are recovered as special cases of Theorem 3,  as shown in Section 5.

The classical mechanical Wakimoto realization  (\ref{11}) 
is valid with respect to arbitrary coordinates
$q_\alpha$ on the group $G_-$. 
However, in order to implement the quantization  
we will first choose  certain special coordinates on $G_-$, 
which we call 
`upper triangular coordinates'. 
Upper triangular coordinates are special `polynomial coordinates'
whose main examples are the  graded exponential 
coordinates  associated to a homogeneous basis $V_\alpha$ of $\G_-$ by
$q\mapsto g_-(q)=\exp\left(\sum_\alpha q_\alpha V_\alpha\right)$.
The polynomiality  of the coordinates    
ensures that the components of $I$ are given by polynomial
expressions, which can be easily rendered  well-defined at the
quantum mechanical level by normal ordering.
The upper triangularity property, defined in Section 3,
is a technical assumption  on polynomial coordinates that  
simplifies the computations.

To  also implement the quantization in polynomial coordinates that are 
not necessarily upper triangular,
in Section 4  we shall investigate  quantum coordinate transformations. 
Two systems of coordinates $q_\alpha$ and $Q_\alpha$ 
on $G_-$ are `polynomially equivalent' if both the coordinate
transformation $q_\alpha(Q)$ and the inverse change of coordinates 
$Q_\alpha(q)$ are given by polynomials.
We shall show that the quantization of the classical Wakimoto realization
can be performed in every system of coordinates which is polynomially
equivalent to an upper triangular system, and all those quantum
mechanical Wakimoto realizations are equivalent  
since they are related by quantum coordinate transformations.

The so-called screening charges play a crucial role in the  applications
of the Wakimoto realization \cite{FF,BF,BMP,ATY,Ku}.
Screening charges are operators  commuting  with the Wakimoto
current that can be used to build resolutions of irreducible 
highest weight representations of the affine algebra 
and to construct chiral primary fields. 
In Section 6, we will find the explicit  form of the screening
charges for each generalized Wakimoto realization, extending 
the result previously known in the principal case 
\cite{BMP,FF,F1,Ku}.

\medskip
\noindent
{\bf Lie algebraic conventions}.
For later reference, 
we here collect some notions of Lie theory  (see e.g.~\cite{GOV}).
Let the step operators
$E_{\pm \alpha_l}$ and the Cartan elements $H_{\alpha_l}$,
associated  with the simple roots 
$\alpha_l$ for $l=1,\ldots, r={\rm rank}(\G)$, 
be the  Chevalley generators of the complex simple Lie algebra $\G$.
For any $l=1,\ldots, r$,  choose an integer
$n_l\in \{ 0,1\}$ and determine the unique Cartan element $H$ for which 
$[H, E_{\pm \alpha_l}]=\pm n_l  E_{\pm \alpha_l}$.
The eigenspaces of $H$ in the adjoint representation 
define a ${\bf Z}$-gradation of $\G$, 
\be
\G =\oplus_m\,  \G_m
\qquad
[\G_m, \G_n] \subset \G_{m+n}
\qquad\hbox{with}\quad
\G_m\equiv \{\, X\in \G\,\vert\, [H, X] = mX\,\}.
\label{grading}\ee 
Denoting  the subspaces of positive/negative grades  by $\G_\pm$,
we obtain the  decomposition 
\be 
\G=\G_- + \G_0 + \G_+.
\label{triang}\ee 
We let $G$ denote a connected complex Lie group 
whose Lie algebra is $\G$.
In terms of  the connected subgroups $G_{0}, G_\pm\subset G$  
corresponding to the subalgebras $\G_{0},\G_\pm$, 
we  have the dense  open submanifold 
$%\be
\check G\equiv \{\, g=g_+g_0 g_-\,\vert \, g_{0}\in G_{0},\,
g_\pm\in G_\pm\,\}
\subset G$ %\label{gauss}\ee
of `Gauss decomposable' elements.
In fact, $\check G$ equals to  $G_+\times G_0 \times G_-$ as a  manifold
since the decomposition of any $g\in \check G$ is unique.
The standard parabolic  subalgebra ${\cal P}\subset \G$ associated with 
the fixed set of integers $n_l$, 
and the corresponding parabolic subgroup $P\subset G$ are given by  
the semidirect products 
\be
{\cal P}=(\G_0+\G_+),
\qquad \quad
P=G_0G_+.
\ee
Any parabolic subalgebra is conjugate to a unique standard one.
For $X, Y\in \G$, we shall denote an  invariant scalar product 
$\langle X,Y\rangle$ simply as $\tr(XY)$ 
as if a matrix representation of $\G$ was chosen. 
Similarly,  we denote say ${\rm Ad\,} g (X)$ as
$gXg^{-1}$  for any $g\in G$, $X\in \G$.
This notation is used purely for convenience, 
a choice of representation is never needed below.
In the principal case, for which $n_l=1$ $\forall l$, 
 $\G_\pm$ are the subalgebras generated by
the positive/negative roots, and 
$\G_0$ is the Cartan subalgebra $\cal H$.
In the general case, the Lie algebra ${{\cal G}}_0$ 
can  be decomposed  into an abelian 
factor, say  $\G_0^0$,
and simple factors, say $\G_0^i$ for $i>0$,  that are orthogonal 
with respect to $\tr$,
\be {\cal G}_0 = \oplus_{i\geq 0}\, {\cal G}_0^i.  
\label{reductive}\ee
If $\psi$ and $\psi_i$ are long roots 
of $\G$ and  of $\G_0^i$ for $i>0$, then the dual Coxeter numbers
of $\G$ and of $\G_0^i$ are respectively given by 
\be
h^*= {c_2(\G)\over \vert\psi\vert^2}\quad\hbox{and}\quad   
h^*_i = {c_2(\G_0^i)\over \vert\psi_i\vert^2} \quad\hbox{for}\quad  i>0.
\label{Cox}\ee
The  scalar product of the roots is defined by identifying 
${\cal H}^*$  with ${\cal H}$ using 
the restriction of the scalar product $\tr$ of $\G$ to $\cal H$.
The quadratic Casimir $c_2(\G)$ of $\G$ is defined by
$\eta^{ab} [T_a,[T_b,Y]] = c_2(\G) Y $
where $Y\in\G$, $\eta^{ab}$ is the inverse of $\eta_{ab}=\tr(T_aT_b)$
for a basis $T_a$ of $\G$; 
and $c_2(\G_0^i)$  is defined  analogously
with the aid of the restriction of the scalar product $\tr$ to $\G_0^i$.

Finally, we have  $\widetilde{A}\equiv C^\infty(S^1,A)$ for 
any  Lie group or Lie algebra $A$. 
The dual space to  $\G$, $\G_0$, $\G_+$ will be respectively
identified with $\G$, $\G_0$, $\G_-$ by means of the scalar 
product, and the analogous  identifications  will be used
for the corresponding loop algebras.

%\newpage
 
\section{Classical Wakimoto realization}
\setcounter{equation}{0}

\def\M{{\cal M}}

Below we derive Poisson bracket (PB)  
realizations of $\widehat{\G}_K$,
 the central extension of the loop algebra 
$\widetilde{\G}=C^\infty(S^1,\G)$, 
by means of Hamiltonian (Marsden-Weinstein)
symmetry reduction \cite{AM} 
of the WZNW model \cite{Wi} based on $G$.
Specifically, the Wakimoto current $I(q,p,j_0)$ in (\ref{11}) 
will appear as one of the affine Kac-Moody 
currents of the WZNW model that survives the reduction, while 
$q_\alpha(\si)$, $p^\alpha(\si)$, $j_0(\si)$ will be interpreted as
coordinates on an open submanifold (the `big cell') 
 in the reduced phase space.
The reduced phase space will not be 
described globally, since this is not needed for the  purposes
of this paper.
%(but see the last footnote)

We take  the  phase space, $\M$,  of the WZNW model to be  
the holomorphic cotangent bundle of the loop group $\widetilde G$, 
realized as  
$$%\be
\M\equiv T^* {\widetilde G}= \{\, (g,J)\,\vert\,
g\in \widetilde{G},\,\,\, J\in \widetilde{\G}\,\}.
$$%\label{WZphase}\ee
The phase space $\M$ is equipped with the symplectic form 
(see e.g.~\cite{HK})
$$%\be
\omega= \int_{S^1} d\, {\rm Tr}\left( J d g g^{-1}\right) 
+ K\int_{S^1}
 {\rm Tr}\left(d g g^{-1}\right) \left(d g g^{-1}\right)' 
$$%\label{WZsymp}\ee
with some $K\in {\bf C}$, which yields the  fundamental PBs 
\begin{eqnarray*}
\{\,\t (T_a\, J)(\si) \,, \, \t (T_b\, J)(\bsi) \,\}_{\rm WZ}
     &=&\t ([T_a,T_b]\, J)(\si) \delta
                +K\, \t (T_a\, T_b) \, \delta',\nonumber\\
\{\,\t (T_a\, J)(\si) \,, \, g(\bsi) \,\}_{\rm WZ}
     &=& - T_a g(\si) \,\delta,
\qquad \delta=\d(\si-\bsi),
\end{eqnarray*}
where $0\leq \sigma\leq 2\pi$ parametrizes $S^1$ and $T_a$ is 
a basis of $\G$.
As a consequence,  
$$%\be
I\equiv  - g^{-1} J g + K g^{-1} g^\prime 
$$%\label{I}\ee
satisfies the PBs 
\be
\{\,\t (T_a\, I)(\si) \,, \, \t (T_b\, I)(\bsi) \,\}_{\rm WZ}
     =\t ([T_a,T_b]\, I)(\si) \delta
                -K\, \t (T_a\, T_b) \, \delta'.
\label{IPB}\ee
The affine Kac-Moody currents 
$J$ and $I$ are the generators (momentum maps) 
for two commuting Hamiltonian actions
of $\widetilde G$ on $T^*\widetilde{G}$.
The action generated  by $J$  
is given by 
$$%\be
L_{\gamma }: (g,J)\mapsto  (\gamma g, \gamma J \gamma^{-1}
+K \gamma' \gamma^{-1})
\qquad\gamma \in \widetilde G.
$$%\label{Lact}\ee
The action generated by $I$ is written  as
$$%\be
R_{\gamma}: (g,J)\mapsto (g \gamma^{-1},J)
\qquad\gamma \in \widetilde G.
$$%\label{Ract}\ee
The action $L_\gamma$ leaves $I$ is invariant,
while $R_\gamma$ preserves $J$ and transforms $I$  
according to the co-adjoint action
\be
R_\gamma: I\mapsto 
\gamma I\gamma^{-1} 
-K \gamma' \gamma^{-1}\,.
\label{Rcoadj}\ee

Now we define a symmetry reduction using the action of 
$\widetilde{P}\subset \widetilde{G}$ given by $L_\gamma$ for
$\gamma \in \widetilde{P}$.
The infinitesimal generators of this symmetry are the components
of the momentum map that maps  $(g,J)\in \M$
to $(J_0+J_-)\in \widetilde{\cal P}^*=
(\widetilde{\G}_0+\widetilde{\G}_-)$, 
where we decompose $J=(J_- + J_0 + J_+)$ using (\ref{triang}).
The reduction is defined by imposing the constraints
$$%\be
J_0 = \mu_0,
\qquad 
J_-=0
$$%\label{constr}\ee
with an arbitrary  $\mu_0 \in \widetilde{\G}_0$.
The corresponding reduced phase space is the space of orbits 
$\M_{\rm red}= \widetilde{P}(\mu_0)\backslash \M_c$,
where $\M_c\subset \M$ is the constrained manifold and 
$\widetilde{P}(\mu_0)$ is the subgroup of $\widetilde{P}$ whose
action preserves the constraints.
Using $\widetilde{G}_{0}\equiv C^\infty(S^1, G_{0})$,
$\widetilde{G}_{\pm}\equiv C^\infty(S^1, G_{\pm})$,
it is easy to see that 
$$%\be
\widetilde{P}(\mu_0)=\widetilde{G}_0(\mu_0) \widetilde{G}_+ 
\qquad\hbox{with}\qquad
\widetilde{G}_0(\mu_0)\equiv \{ g_0 \in \widetilde{G}_0\,\vert\,
g_0 \mu_0 g_0^{-1} + K g_0' g_0^{-1}=\mu_0\,\}.
$$%\label{isotr}\ee
We are only interested in the big cell $\check \M_c\subset \M_c$
where $g$ is Gauss decomposable, which is a property respected  by 
the action of $\widetilde{P}$.
Our immediate aim is to characterize the manifold   
$$%\be
\check \M_{\rm red}= \widetilde{P}(\mu_0)\backslash \check{\M}_c\,,
\qquad 
\check{\M}_c=\{ (g_+ g_0 g_-, J_+ + \mu_0)\,\vert\, 
g_{0}\in \widetilde{G}_{0},\,\,\,
g_{\pm}\in \widetilde{G}_{\pm},\,\, J_+\in \widetilde{\G}_+\,\},
$$%\label{Mred}\ee
and to find the symplectic form  $\omega_{\rm red}$
on $\check \M_{\rm red}$ defined by the canonical map
$\eta: \check{\M}_c \rightarrow \check{\M}_{\rm red}$ as
$\eta^* \omega_{\rm red} = \omega\vert \check{\M}_c$.
For this we notice that a complete set of 
$\widetilde{P}(\mu_0)$ invariant functions on  $\check{\M}_c$ is given by 
\be
g_-,\quad j\equiv  g_0^{-1} \left( g_+^{-1}\left(J_+ +\mu_0\right)g_+-\mu_0 
-Kg_+^{-1} g_+' \right)g_0,
\quad
j_0\equiv  - g_0^{-1} \mu_0 g_0 +K g_0^{-1}g_0'.
\label{invars}\ee
The induced mapping 
$$%\be
(g_-, j, j_0): \check\M_{\rm red} \rightarrow \widetilde{G}_- \times 
\widetilde{\G}_+
\times \widetilde{G}_0(\mu_0)\backslash \widetilde{G}_0,
$$%\ee 
is in fact {\em one-to-one}, and hence $(g_-,j,j_0)$ may be regarded as 
coordinates on the reduced phase space.
Next, one verifies  that 
\be
\omega\vert \check \M_c= \int_{S^1} d\, {\rm Tr}
\left( j d g_- g_-^{-1}\right) 
+ \int_{S^1} d\, {\rm Tr}\left( \mu_0  d g_0 g_0^{-1}\right) 
+ K\int_{S^1}
 {\rm Tr}\left(d g_0 g_0^{-1}\right) \left(d g_0 g_0^{-1}\right)'. 
\label{onMc}\ee
The first term is then recognized to be the canonical 
symplectic form on $T^* \widetilde{G}_-= 
\widetilde{G}_- \times \widetilde{\G}_+$, 
where the identification 
is made using right tranlations to trivialize 
$T^* \widetilde{G}_-$ and  
$\widetilde{\G}_-^*=\widetilde{\G}_+$.
To interpret the other two terms, we note that the coset space 
$\widetilde{G}_0(\mu_0)\backslash \widetilde{G}_0$ can be realized 
as the orbit, ${\cal O}^0_{-K}(-\mu_0)$, 
of $\widetilde{G}_0$ through the point $j_0=-\mu_0$ 
with respect to the following
action of $\widetilde{G}_0$ on $\widetilde{\G}_0$:
\be
R^0_{g_0}: j_0\mapsto 
g_0 j_0 g_0^{-1} 
-K g_0' g_0^{-1}\,
\qquad g_0\in \widetilde{G}_0,\,\, j_0\in \widetilde{\G}_0.
\label{Rcoadj0}\ee
This is the co-adjoint action in (\ref{Rcoadj}) applied to 
$\widetilde{G}_0$, and thus it preserves the PB
\be
\{\,\t (Y_i\, j_0(\si)) \,, \, \t (Y_l\, j_0(\bsi)) \,\}
     =\t ([Y_i,Y_l]\, j_0(\si)) \delta
                -K\, \t (Y_i\, Y_l) \, \delta',
\label{j0PB}\ee
where $Y_i$ is a basis of $\G_0$ and 
%, abusing the notation,
$j_0$ now denotes  a coordinate on $\widetilde{\G}_0$.
Actually, we  can identify the sum of the second and
third terms in (\ref{onMc}) to be just the canonical symplectic form 
on the co-adjoint orbit 
${\cal O}^0_{-K}(-\mu_0)=\widetilde{G}_0(\mu_0)\backslash \widetilde{G}_0$
in terms of the redundant coordinate $g_0$.
In order to confirm this formula of the canonical symplectic form,
it is enough to remark that in the full WZNW model $I$ runs over 
the co-adjoint orbit of (the central extension of) $\widetilde{G}$ 
through $-\mu$ for any fixed $J=\mu$, and thus the 
formula for the symplectic form on this orbit follows from
the reduction of the WZNW model defined by the constraint $J=\mu$.
Then apply this remark to $\widetilde{G}_0$. 
The outcome of the foregoing  analysis is summarized as follows.

\medskip\noindent
{\bf Theorem 1.} {\em  
The big cell $\check{\M}_{\rm red}$ 
of the reduced phase space can be identified with the manifold
$T^* \widetilde G_- \,\times\, {\cal O}^0_{-K}(-\mu_0)$ where 
${\cal O}^0_{-K}(-\mu_0)=\widetilde{G}_0(\mu_0)\backslash \widetilde{G}_0$ 
is the orbit of $\widetilde{G}_0$ through 
$-\mu_0$ with respect to the co-adjoint action in (\ref{Rcoadj0}).
The  symplectic form on $\check{\M}_{\rm red}$
defined by $\omega\vert \check\M_c$ in (\ref{onMc}) coincides with the 
canonical symplectic form on this product manifold.
}
\medskip

The Poisson brackets  of the coordinate functions $(g_-, j, j_0)$
determined by the symplectic form on $\check{\M}_{\rm red}$ 
(which can be thought of as Dirac brackets) are given by (\ref{j0PB})
together with 
\begin{eqnarray}
\{\,\t (V_\alpha\, j)(\si) \,, \, \t (V_\beta\, j)(\bsi) \,\}
     &=& \t ([V_\alpha,V_\beta]\, j)(\si) \,\d,
\nonumber \\
\{\,\t (V_\alpha\, j)(\si) \,, \, g_-(\bsi) \,\}
     &=& - V_\alpha\, g_-(\si) \,\d,
\label{jPB}\end{eqnarray}
where $V_\alpha$ is a basis of $\G_-$.
Denote by $V^\beta$ the dual basis of $\G_+$, 
$\tr(V^\beta V_\alpha)=\delta_\alpha^\beta$.
Let $q_\alpha$ be some global, holomorphic coordinates on $G_-$.
Define 
\be 
N^{\alpha\beta}(q)\equiv  
\tr(V^\beta {\pa g_- \over \pa q_\alpha} g_-^{-1}).
\label{N}\ee
Then\footnote{\ninerm 
To compare with (\ref{9}), 
we have  ${\cal F}_{\alpha\beta}=N^{-1}_{\alpha\beta}$.} 
\be
j(q,p)=N^{-1}_{\alpha\beta}(q) p^\beta V^\alpha 
\label{jform}\label{N-1}\ee
in terms of the canonical coordinates $q_\alpha(\sigma)$, 
$p^\beta(\sigma)$ on $T^*\widetilde{G}_-$.
Indeed, 
$$%\be
\int_{S^1} d\, {\rm Tr}
\left( j d g_- g_-^{-1}\right) =\int_{S^1} d 
\left( p^\alpha dq_\alpha\right)\,,
$$%\ee
and therefore 
\be
\{ q_\alpha(\sigma), p^\beta(\bar\sigma)\} =\delta_\alpha^\beta 
\delta(\sigma-\bar\sigma).
\label{pqPB}\ee
The classical Wakimoto realization is given by the statement:

\medskip\noindent
{\bf Theorem 2.} {\em As a consequence of the  PBs of  
$j_0$ in  (\ref{j0PB}) and $q_\alpha$, $p^\beta$  in (\ref{pqPB}),
where (\ref{pqPB}) is equivalent to (\ref{jPB}) by 
means of (\ref{jform}), 
 the classical Wakimoto current 
\be
I(q,p,j_0) =I(g_-(q), j(q,p), j_0)=
g_-^{-1} ( -j + j_0) g_-  + K g_-^{-1} g_-'
\label{affwak}\ee
satisfies 
\be
\{\,\t (T_a\, I)(\si) \,, \, \t (T_b\, I)(\bsi) \,\}
     =\t ([T_a,T_b]\, I)(\si) \delta
                -K\, \t (T_a\, T_b) \, \delta'.
\label{KMPB}\ee
The affine-Sugawara stress-energy tensor is quadratic 
in the free field constituents,  
\be \label{1.11}
{1\over 2K} \t ( I^2) = {1\over 2K}\t (j_0^2) - 
\t (j g_-' g^{-1}_-)={1\over 2K}\t (j_0^2) 
- \sum_\alpha p^\alpha q_\alpha'.
\ee
}
\medskip
\noindent{\em Proof.}
With the $\widetilde{P}(\mu_0)$ invariants 
 in (\ref{invars}) on $\check{\M}_c$, we have 
$$
I = -(g_+ g_0 g_-)^{-1} (J_+ +\mu_0) (g_+ g_0 g_-)
+ K (g_+ g_0 g_-)^{-1} (g_+ g_0 g_-)'=
g_-^{-1} ( -j + j_0) g_- + K g_-^{-1} g_-' .
$$
This implies the first statement since the PBs of $I$ in 
(\ref{IPB}) survive the reduction as $I$
is invariant under the symmetry group $\widetilde P$. 
 The second statement is easily  verified. {\em Q.E.D.}

\medskip\noindent
{\em  Remark 1.} 
The validity of  Theorem 2 does not require
$j_0$ to be restricted to a co-adjoint orbit.
This is clear, for example, from the fact
that the orbit appearing  in Theorem 1 is 
through an {\it arbitrarily}  chosen element $-\mu_0$. 
\smallskip

%\newpage

It is natural to ask 
what the result would be if one substituted 
a Wakimoto realization of the $\G_0$-valued current $j_0$ into the 
$\G$-valued Wakimoto current $I(g_-,j,j_0)$ in (\ref{affwak}).
In fact, the resulting $\G$-valued 
current will have  the form (\ref{affwak}) with respect to 
another parabolic subalgebra of $\G$.
Below  we explain  this `composition property' 
of the classical Wakimoto realization.

We need to introduce some notations to verify the composition property. 
Suppose that we are given the parabolic subalgebra 
${\cal P}= (\G_0 + \G_+)$ associated with the triangular
decomposition $\G=\G_- + \G_0 + \G_+$.
Then  consider a parabolic subalgebra ${\cal P}_0\subset \G_0$,
where  ${\cal P}_0= (\G_{0,0} +\G_{0,+})$ in terms of a triangular  
decomposition $\G_{0}= \G_{0,-} + \G_{0,0} + \G_{0,+}$, which 
is defined by the 
signs of the 
eigenvalues of some real-semisimple element $H_0\in \G_0$.
On account of  standard Lie algebraic results \cite{GOV}, 
${\cal P}^{\rm c} \equiv  ({\cal P}_0  + \G_+)\subset \G$
is again   a parabolic subalgebra.
Here the superscript `c' is our mnemonic for `composite'.
We have  the new triangular 
decomposition
$$
\G= \G_-^{\rm c} + \G_{0,0} + \G_+^{\rm c}
\qquad\hbox{with}\qquad
\G_{\pm}^{\rm c}\equiv  \G_{0, \pm } + \G_{\pm},
$$
and the subalgebras  
$\G_{0,0}$, $\G_\pm^c$
allow us to further identify ${\cal P}^c$  in the form 
$$
{\cal P}^{\rm c} = {\cal P}_0  + \G_+ =  \G_{0,0} + \G_{0,+} + \G_+ =
\G_{0,0} + \G_+^c.
$$
Notice that  $\G_\pm^{\rm c}$ are
semidirect sums since $[\G_{0,\pm}\,,\, \G_{\pm}] \subset \G_{\pm}$.
In correspondence with all these Lie subalgebras, we also introduce
the respective connected subgroups $G_{0,0}$ and $G_{0,\pm}$ of $G_0$,
and the semidirect product groups 
$G_{\pm}^{\rm c}=G_{0,\pm} G_\pm$, where $G_\pm$ and $G_0$ are 
the subgroups of $G$ associated with the original 
triangular decomposition.

Armed with these notations, we can now describe the `sub-Wakimoto' 
realization
of $j_0$ in a form analogous (\ref{affwak}), namely  
$$
j_0 = n_-^{-1} ( - i + i_0 ) n_- + K n_-^{-1} n_-'
$$
where $i_0$ is $\G_{0,0}$-valued, 
$i$ is $\G_{0,+}$-valued, and $n_-$ is $G_{0,-}$-valued.
Substituting this into (\ref{affwak}) we then find the identity
$$
g_-^{-1} ( -j + j_0) g_-  + K g_-^{-1} g_-'=
(g_-^c)^{-1} ( -j^c + j_0^c) g_-^c  + K (g_-^c)^{-1} (g_-^c)'
$$
with the composite objects 
$$
g_-^c = n_- g_-,
\qquad
j^c = i+ n_- j n_-^{-1},
\qquad
j_0^c = i_0
$$
that belong to $G_-$, $\G_+^c$ and $\G_{0,0}$, respectively.
This identity expresses 
the composition property of the Wakimoto 
realization, which  may be symbolically written as 
\be
I(g_-, j, j_0(n_-, i, i_0)) = I( g_-^c, j^c, j_0^c).
\label{comp}\ee
After postulating   the Poisson brackets of the 
constituents variables  $g_-$, $j$ and $n_-$, $i$, $i_0$,
the Poisson brackets of the
composite variables  $g_-^c$, $j^c$, $j_0^c$ are readily checked to be 
the correct ones that one requires  for the Wakimoto realization based 
on ${\cal P}^c$. 

The coordinates on $G_-^{\rm c}$ that are naturally
associated with the composition property just exhibited 
are obtained as unions of independent coordinates on $G_-$ 
and on $G_{0,-}$.
Using such coordinates, 
the composition property in principle allows one  
to produce the principal Wakimoto realization of $I\in \G$
by proceeding through a chain of non-principal  Wakimoto realizations
according to a partial  ordering of parabolic subalgebras.
%%These  remarks will be illustrated later in examples.

%\newpage

\section{Quantization of the classical Wakimoto realization}
\setcounter{equation}{0}

Our goal now is to derive a quantum counterpart of the
classical Wakimoto realization  (\ref{affwak}). As this classical
realization was derived by means of a Hamiltonian reduction, there seem
a priori to be two ways to quantize it.
The first possibility would require us to write down a 
quantization of the
phase space of the WZNW model and
subsequently to implement a quantum Hamiltonian
reduction.
Although it might be very interesting to pursue this line of thought
further, it seems to be rather cumbersome, and in addition in the case
at hand it turns out to be relatively easy to directly quantize the
classical Wakimoto realization. Therefore we will restrict ourselves to
the latter method. 

Since the classical Wakimoto realization expresses the currents of $\G$
in terms
of currents in ${\cal G}_0$ and a set of coordinates
and momenta that constitute 
the cotangent bundle $T^*{\widetilde G}_-$,  our philosophy will
be to first quantize these objects by postulating 
OPEs  for their generators, and subsequently to write down
a normal ordered version of (\ref{affwak}) in terms of these generators.
The hard work lies  in verifying that the currents defined in this way
indeed satisfy the OPEs of the  affine Lie algebra based on ${\cal G}$.
This requirement will in addition  fix the 
ambiguities that one has to deal with in normal ordering
(\ref{affwak}).

Fixing a basis $T_a$ of $\G$, 
 the OPEs corresponding 
to (\ref{KMPB}) should  read as 
\be
\tr (T_a \underhook{I)(z) \tr(T_b} I)(w)  =
\frac{K \tr( T_a T_b)}{(z-w)^2} +
\frac{\tr( [T_a,T_b] I)(w)}{z-w}.
\label{requi}
\ee
Replacing (\ref{KMPB}) with (\ref{requi})
amounts to replacing the PBs of the Fourier modes of the current   with  
corresponding commutators, as is well-known.
Naturally, 
the OPEs of the constituent coordinate and momentum fields are  
declared to be
\be 
p^{\al}\underhook{(z) q_{\bt}}(w) = \frac{-\delta^{\al}_{\bt} }{z-w}.
\label{pqOPE}\ee
Decomposing  the $\G_0$-valued current $j_0$ as $j_0
=\sum_{i\geq 0} j_0^i$
according to (\ref{reductive}),
we postulate the OPEs  of the current $j_0^i$ in ${\cal G}_0^i$  as   
\be
\tr (\pi_0^i(T_a) \underhook{ j_0^i)(z) \tr(\pi_0^i(T_b) } j_0^i)(w)  =
\frac{K_0^i \tr( \pi_0^i(T_a) \pi_0^i(T_b))}{(z-w)^2} +
\frac{\tr( [\pi_0^i(T_a),\pi_0^i(T_b)] j_0^i)(w)}{z-w}
\label{j0i}\ee
where $\pi_0^i:{\cal G} \rightarrow {\cal G}_0^i$ is the orthogonal 
projection onto ${\cal G}_0^i$.
All other OPEs of the constituents are regular. 
Note that we have taken the central
extension $K_0^i$ of $j_0^i$ to be a free parameter,
to be determined from requiring (\ref{requi}), 
and that the properly normalized {\em level} parameters of $I$ 
and  $j_0^i$ 
(which are integers in a  unitary highest weight representation)
are respectively given by  
$$%\be
k\equiv {2K\over \vert \psi\vert^2}
\qquad\hbox{and}\quad
 k_0^i\equiv {2 K_0^i\over \vert \psi_i\vert^2}\quad\hbox{for}\quad i>0.
$$%\label{level}\ee

Notice now that the  classical Wakimoto current in (\ref{affwak}) 
is linear in the $p^\al$ and in $j_0$, but  could contain 
arbitrary functions 
of the $q_\alpha$ if the coordinates  were not chosen with care. 
However, we here only  wish to deal with objects that are polynomial 
in the 
basic quantum fields, since those are easily defined in chiral conformal 
field theory 
(which is the same as the theory of vertex algebras,
see e.g.\ \cite{VA})  by 
normal ordering.
Below we will define a class of coordinates 
on $G_-$, the so-called `upper triangular coordinates', in which
the quantum Wakimoto current will be polynomial.
The computations will also simplify considerably in these coordinates.

Fortunately, the  only ordering problem in (\ref{affwak}) arises from 
the term $-g_-^{-1} j g_- = -g_-^{-1} N^{-1}_{\alpha\beta}(q) p^{\beta}
V^{\alpha} g_-$   for which we have
to choose where to put the momenta $p^{\beta}$.
We  now choose
to put them on the left, and  replace any  classical
object $pf(q)$ by the normal ordered object $(p(f(q)))$. 
{}From the 
OPE\footnote{\ninerm The notations 
${\partial^\alpha f}(q)\equiv 
{\partial f(q)\over \partial q_\alpha}$ and  $(\partial F)(z)\equiv
{\partial F(z) \over \partial z}$ are used 
for functions $f$ of $q$ and $F$ of $z$ from now on.}
$$%\be 
p^{\alpha}\underhook{(z) f(q(}w)) = \frac{-\partial^{\alpha} f}{z-w}
$$%\ee
we see 
that $\left([p^{\alpha},f(q)]\right)=-\partial \partial^{\alpha} f(q)
=-\partial^{\beta} \partial^{\alpha} f(q) \dif q_{\beta}$. Hence, 
the difference between two normal orderings of the classical object
$pf(q)$ will always be of the form $\Omega^{\beta}(q) \partial q_{\beta}$
for some function $\Omega^{\beta}$, and we should allow for an additional
term of this type in  quantizing (\ref{affwak}). Altogether this leads
to the following proposal for the quantum Wakimoto current:
\be \label{qwak}
I = -\left(p^{\bt}(N^{-1}_{\al\bt} g_-^{-1} V^{\al} g_-)\right)
+ g_-^{-1} j_0 g_- +  K g_-^{-1} \dif g_- +
g_-^{-1} \Omega^{\beta} g_- \dif q_{\beta}. 
\ee
Our main result will be  to give the  explicit form of 
the last term, which  represents a quantum correction  due to different
normal orderings.
The function $\Omega^{\beta}(q)$ is  ${\cal G}$-valued and
we inserted some factors of $g_-$ around it for convenience.

To  define  the special  coordinates
in which our formula  for $\Omega^\beta$  will be valid,
we first introduce the matrix  $R^b_a(g_-)$ by 
\be 
g_- T_a g_-^{-1}\equiv  R^b_a(g_-)T_b
\qquad g_-\in G_-.
\label{Ad}\ee 
\noindent
{\bf Definition (polynomial coordinates).} 
We  call a  system of global, holomorphic coordinates
$q_\alpha$ on $G_-$  {\em polynomial} if 
$R_a^b(g_-(q))$ is given by a  polynomial  of the  coordinates. 
\smallskip

The fact that   $\det R=1$, which follows from the invariance of $\tr$
and from the fact that $G_-$ is topologically trivial,  
shows that  $R^b_a(g_-^{-1})$  is  also polynomial in the $q_\alpha$.
Furthermore, since one can evaluate $N^{\alpha\beta}(q)$ 
in (\ref{N}) using  the adjoint representation,
the definition implies that $N^{\alpha\beta}(q)$ is a polynomial, too.
The determinant of $N^{\alpha\beta}(q)$
is a nonwhere vanishing complex polynomial and 
 must  therefore be a constant, so that 
$N^{-1}_{\alpha\beta}(q)$ is  a polynomial as well.

Let  ${\rm deg}$  denote the ${\bf Z}$-gradation with respect to 
which the decomposition (\ref{triang}) was 
made\footnote{The  standard  `grading operator'  $H$ in
(\ref{grading}), $[H, E_{\pm \alpha_l}]=\pm n_l E_{\pm \alpha_l}$,
can  be replaced  with any  $H'$ for which 
$[H', E_{\pm \alpha_l}]=\pm n'_l E_{\pm \alpha_l}$ in such a way that 
$n'_l=0$ whenever $n_l=0$ and $n'_l n_l >0$ otherwise. 
All results remain  true if one uses  
the subsequent definition relative to any such gradation.},  
assume that the 
basis elements $V_{\alpha}$ of $\G_-$  
have  well-defined degree, and set 
$d_\alpha \equiv  -{\rm deg}(V_{\al}) = {\rm deg}(V^{\al})>0$.
For  polynomial coordinates $q_\alpha$  on $G_-$,  
let us assign degree $d_\alpha$ to $q_\alpha$.

\smallskip
\noindent
{\bf Definition (upper triangular coordinates).}
We call a system  of polynomial coordinates on $G_-$ 
{\em  upper triangular\,} if 
$N^{\alpha\beta}(q)$ is given by a homogeneous polynomial of degree 
$(d_\beta-d_\alpha)$ with respect to the above  assignment 
of the degree. 
 
\smallskip
\noindent 
{\em Remark 2.}
In upper triangular coordinates
$N^{\al\bt}(q)$ obviously vanishes  
unless $d_{\bt} \geq d_{\al}$, which  explains  the name
and implies that 
$N^{-1}_{\al\bt}(q)$ is  also a polynomial of  degree 
$(d_{\bt}-d_{\al})$.
Thus the definition ensures that 
the vector field $N^{-1}_{\al\bt}(q) {\dif \over \dif q_\beta}$, 
which generates the one-parameter 
group $g_- \mapsto e^{tV_\alpha} g_-$, has 
the degree $-d_\alpha$ of $V_\alpha$ when acting on polynomials in the 
coordinates. 
\smallskip

\noindent
{\em Examples.}
The most obvious examples of upper triangular coordinates are the
`graded exponential coordinates', given by $g_-(q)=\exp(
\sum_{\alpha} q_{\alpha} V_{\alpha} )$. One can also take products
of graded exponential coordinates, by distributing the 
set $\{ V_{\al} \}$
over disjoint subsets $S_{\cal I}$ and taking
$
g_-(q) = 
\prod_{\cal I} \exp( \sum_{\alpha \in S_{\cal I}} q_{\al} V_{\al} ).
$
If $G=SL(n)$,  
there are even simpler coordinates where the
$q_{\alpha}$ are matrix elements, namely $g_-(q)={\bf 1}_n+\sum_{\alpha}
q_{\alpha} V_{\alpha}$ with $(V_\alpha)_{lk}=\delta_{il}\delta_{jk}$ 
for some $i>j$.
To check the upper triangularity property in these examples,
it is useful to think of $g_-(q)$ as a polynomial
in the $q_\alpha$ and the $V_\alpha$ which are 
declared to have degrees $d_\alpha$ and $-d_\alpha$, respectively.
Then  $g_-(q)$ has  `total degree' zero, and 
$N^{\alpha\beta}(q) V_\beta$ has total degree $-d_\alpha$, 
implying that $N^{\alpha\beta}(q)$ has degree $(d_\beta-d_\alpha)$.
For the same reason, in our examples not only $N^{\alpha \beta}(q)$,
but actually also the matrix $R_a^b(q)$ in (\ref{Ad}) is given by 
homogeneous polynomials. 
Namely, if $[H, T_a] = \tau_a T_a$ 
with the grading operator $H$ in (\ref{grading}), 
then $ R_a^b (q)$ has degree  $(\tau_a - \tau_b)$.
It follows that the vector field that generates  the action of $T_a$
on $G_- \subset P\backslash G$, induced from right multiplication,
has degree $\tau_a$ when acting on polynomials in the  coordinates.

Now we are ready to state the  main result of this section: 

\noindent
\medskip\noindent
{\bf Theorem 3.} 
{\em  Given a system  of upper triangular coordinates $q_{\al}$ on 
$G_-$, 
the current
${I}$ defined in (\ref{qwak}) satisfies the
OPE given in (\ref{requi}) if 
(i) $2 K_0^0={\vert \psi\vert^2 }(k+h^*)= \vert\psi_i\vert^2 
(k_0^i+h_i^*)$ 
for $i>0$,
and (ii) $\Omega^{\beta}$ is given by the following ${\cal G}_-$-valued
object
\be \label{omega}
\Omega^{\beta} = N^{\lambda\rho} \partial^{\beta} N^{-1}_{\gamma\lambda}
 [V^{\gamma},V_{\rho}].
\ee
Furthermore, $\Omega^{\beta}$ is uniquely determined by (\ref{requi})
up to trivial 
redefinitions of the momenta $p^{\beta}$  in (\ref{qwak}),  
\be \label{predef}
p^{\beta} \rightarrow p^{\beta} + (\partial^{\beta} A^{\gamma} - 
\partial^{\gamma} A^{\beta} ) \partial q_{\gamma} 
\ee
with an arbitrary polynomial $A^{\gamma}(q)$. 
Finally, the  affine-Sugawara stress-energy tensor for the current ${I}$
is equal to a sum consisting  of 
free stress-energy tensors for $p^{\beta},q_{\beta}$,
affine-Sugawara stress-energy tensors for the currents 
$j_0^i$ with values in the simple factors ${\cal G}_0^i\subset \G_0$,
and an improved stress-energy tensor for the current $j_0^0$ 
with values in the  abelian factor $\G_0^0\subset \G_0$, 
\be \label{sugawara}
\frac{1}{2y} \tr ({I}^2)  =
- p^{\beta} \dif q_{\bt} + \frac{1}{2y}\sum_{i\geq 0} \tr(j_0^i j_0^i) 
+\frac{1}{y}\tr((\rho_{{\cal G}} - \rho_{{\cal G}_0} ) \dif j_0^0)
\ee
where  $y\equiv  {1\over 2}\vert \psi\vert^2 (k+h^*)$ and 
$\rho_{{\cal G}} - \rho_{{\cal G}_0} = {1\over 2} 
[V^{\alpha},V_{\alpha}]$
is half the sum of those positive roots of $\G$ that are 
not roots of $\G_0^i$ for any $i>0$.}

\medskip\noindent
{\em Proof.}
The proof of this theorem consists of an explicit calculation
of the OPE of $I$ with itself and of $\tr(I^2)$.
This calculation is relatively straightforward and not very insightful,
so we will not present it in full detail. In it one needs for example 
the relation between the structure constants of ${\cal G}_-$ and
$N^{\al\bt}$, which follows directly from (\ref{N}),
\be \label{struc}
N^{-1}_{\delta \gm} \dif^{\gamma} N^{-1}_{\al\bt} N^{\bt\mu}-
N^{-1}_{\al \gm} \dif^{\gamma} N^{-1}_{\delta \bt} N^{\bt\mu}=
f_{\al\delta}{}^{\mu}. 
\ee
In addition, one can use the fact that the coordinates $q_{\al}$ are
upper triangular to show that many terms appearing in the calculation
actually vanish. To illustrate how the calculations are done, we
compute here only the double pole in the OPE of $\tr(T_a I)(z)$
with $\tr(T_b I)(w)$. Denote by $I_1,\ldots,I_4$ the four terms
on the right hand side of (\ref{qwak}). Then the OPE of
$\tr(T_a I_1)(z)$ with $\tr(T_b I_1)(w)$ has a double pole 
$D_{1,1}(w)/(z-w)^2$ which is completely absent in the classical case,
with
$$%\be
D_{1,1} = - \dif^{\gm} \tr(T_a N^{-1}_{\al\bt} g_-^{-1} V^{\al} g_-)
 \dif^{\bt} \tr(T_b N^{-1}_{\delta\gm} g_-^{-1} V^{\delta} g_-).
$$%\ee
This follows from the OPE
$$%\be
(p^{\al}f\underhook{ (q))(z) (p^{\bt}g(q}))(w) =
\frac{-(\dif^{\bt}f(z) \dif^{\al}g(w))}{(z-w)^2} +
\frac{(p^{\al}(g \dif^{\bt} f)-p^{\bt} (f \dif^{\al} g))(w)}{z-w}.
$$%\ee
Before we continue, we introduce the following notation. Define
\be \tilde{T}_a=g_- T_a g_-^{-1}, \ee
and let $\tilde{T}_a=\tilde{T}_a^- + \tilde{T}_a^0 + \tilde{T}_a^+ $
be the decomposition of $\tilde{T}_a$ in parts  with values in
$\G_+$, $\G_0$, $\G_-$, and $\tilde{T}_a^{0,i}$ the part of
$\tilde{T}_a^0$ with values in $\G_0^i$. The next double pole,
coming from the OPE of $I_2$ with itself, can now be written as
$D_{2,2}(w)/(z-w)^2$ with 
$$%\be
D_{2,2} = \sum_i K_0^i \tr(\tilde{T}_a^{0,i} \tilde{T}_b^{0,i})
$$%\ee
The only further double poles come from the OPE of $I_1$ with
$I_3$ and from the OPE of $I_1$ with $I_4$. These contribute
double poles
$$%\be
D_{1,3} = K \tr(\ttt_a^- \ttt_b^+ + \ttt_a^+ \ttt_b^-)
$$%\ee
and
$$%\be
D_{1,4}= \tr(\ttt_a V^{\al} N^{-1}_{\al\bt})
         \tr(\ttt_b \Omega^{\bt})
        +\tr(\ttt_b V^{\al} N^{-1}_{\al\bt})
         \tr(\ttt_a \Omega^{\bt}).
$$%\ee
According to (\ref{requi}), the sum of all the double poles should be
equal to $K\tr(T_aT_b)$. Subtracting this from the sum of the double
poles yields the equation
$$%\be
D\equiv  D_{1,1}+D_{1,4} + \sum_i (K_0^i-K) \tr( \ttt_a^{0,i} 
\ttt_b^{0,i} )=0.
$$%\ee
More explicitly, $D$ is equal to
\bea \label{deq}
D& = &
-\tr ( [V_{\rho},\ttt_a] V^{\al}) \tr([V_{\al},\ttt_b] V^{\rho})
\nonu & &
-\tr ( [V_{\rho},\ttt_a] V^{\al}) N^{\gamma\rho} N^{-1}_{\al\bt}
 \dif^{\bt} N^{-1}_{\delta\gamma} \tr(\ttt_b V^{\delta})
\nonu & &
-\tr ( \ttt_a V^{\al}) \dif^{\gamma} N^{-1}_{\al\bt}
 N^{\bt\eta} N^{-1}_{\delta\gamma} \tr([V_{\eta},\ttt_b] V^{\delta})
\nonu & &
-\tr(\ttt_a V^{\al} )  \dif^{\gamma} N^{-1}_{\al\bt}
 \dif^{\bt} N^{-1}_{\delta\gamma} \tr(\ttt_b V^{\delta})
\nonu & &
 + \tr(\ttt_a V^{\al} )  N^{-1}_{\al\bt}
 \tr(\ttt_b \Om^{\bt} )
\nonu & &
 +\tr(\ttt_a \Om^{\bt} )
 \tr(\ttt_b V^{\al} )  N^{-1}_{\al\bt}
\nonu & &
+\sum_i (K_0^i-K) \tr (\ttt^{0,i}_a \ttt^{0,i}_b).
\eea
The fourth term in $D$ vanishes identically in upper triangular
coordinates. The first term can be further written as
\begin{eqnarray*}
-\tr ( [V_{\rho},\ttt_a] V^{\al}) \tr([V_{\al},\ttt_b] V^{\rho})
& = & 
-\tr ( [V_{\rho},\ttt^0_a] V^{\al}) \tr([V_{\al},\ttt^0_b] V^{\rho}) 
\nonu
& & 
-\tr ( [V_{\rho},\ttt^-_a] V^{\al}) \tr([V_{\al},\ttt^+_b] V^{\rho}) 
\nonu
& & 
-\tr ( [V_{\rho},\ttt^+_a] V^{\al}) \tr([V_{\al},\ttt^-_b] V^{\rho}) 
\nonu
& = & 
-\tr ( [V_{\rho},\ttt^0_a] V^{\al}) \tr([V_{\al},\ttt^0_b] V^{\rho}) 
\nonu
& & 
-\tr ( \ttt_a V^{\delta}) f_{\rho\delta}{}^{\alpha}
   \tr([V_{\al},\ttt^+_b] V^{\rho}) \nonu
& & 
-\tr ( [V_{\rho},\ttt^+_a] V^{\al}) f_{\alpha\delta}{}^{\rho}
\tr(\ttt^-_b V^{\delta}). 
\end{eqnarray*}
If we now use (\ref{struc}) to replace the structure constants in this
last expression by expressions involving $N$, they conspire with the
second and third line in (\ref{deq}) to yield the following
expression for $D$
\bea
D& = & \label{eq00}
-\tr ( [V_{\rho},\ttt^0_a] V^{\al}) \tr([V_{\al},\ttt^0_b] V^{\rho}) 
\nonu & &
 + \tr(\ttt_a V^{\al} )  N^{-1}_{\al\bt}
 \tr(\ttt_b (\Om^{\bt}-N^{\lambda\rho} \dif^{\beta} 
N^{-1}_{\delta\lambda}
 [V^{\delta},V_{\rho}] ))
\nonu & &
 +\tr(\ttt_a (\Om^{\bt} -N^{\lambda\rho} 
\dif^{\beta} N^{-1}_{\delta\lambda}
 [V^{\delta},V_{\rho}]))
 \tr(\ttt_b V^{\al} )  N^{-1}_{\al\bt}
\nonu & &
+\sum_i (K_0^i-K) \tr (\ttt^{0,i}_a \ttt^{0,i}_b).
\eea
The first and fourth line contain only the projections 
of   $\ttt_a$ and $\ttt_b$ on  $\G_0$, whereas the second and third line 
contain the projections of $\ttt_a$ and $\ttt_b$ 
on  $\G_-$, respectively.
 Therefore the sum of the first and fourth line must
vanish independently of the sum of the second and third line.
The sum of the first and fourth line can be written as
\be \label{eq01}
\sum_i (K^i_0-K)\tr(\ttt^{0,i}_a \ttt^{0,i}_b )
-\frac{1}{2} \tr ( \ttt^0_a [V_{\rho}, [V^{\rho}, \ttt^0_b ] ] )
-\frac{1}{2} \tr ( \ttt^0_a [V^{\rho}, [V_{\rho}, \ttt^0_b ] ] ).
\ee
Using the quadratic  Casimirs $c_2(\G)$,  $c_2(\G_0^i)$ that appear in 
(\ref{Cox}) and $c_2(\G_0^0)=0$,
we find from (\ref{eq01}) that the following identity must hold
$$%\be
\sum_{i\geq 0}\left(K^i_0+{1\over 2}c_2(\G_0^i)-K-
{1\over 2}c_2(\G)\right)
\tr(\ttt^{0,i}_a \ttt^{0,i}_b )=0.
$$%\ee
This implies the relation 
$$%\be 
K^i_0+ {1\over 2} c_2(\G_0^i)= K + {1\over 2} c_2(\G), \qquad
\forall i\geq 0, 
$$%\ee 
which is claimed in Theorem 3. 
The remaining two terms in (\ref{eq00}) have
to vanish separately, and this gives a linear equation for 
$\Omega^{\bt}$.
The general  solution to this linear equation is  given by a particular
solution of the inhomogeneous equation plus the  general solution,
say $\Delta^\beta$, 
to the homogeneous equation. From (\ref{eq00}) we see immediately that
a particular solution 
of the inhomogeneous equation  for $\Omega^{\bt}$   is precisely the
one given in Theorem 3. It remains to analyze the homogeneous 
equation
\be \label{eq02}
  \tr(\ttt_a V^{\al} )  N^{-1}_{\al\bt}
 \tr(\ttt_b \Delta^{\bt}) 
 +\tr(\ttt_a \Delta^{\bt} ) N^{-1}_{\al\bt}
 \tr(\ttt_b V^{\al} )  =0.
\ee
Decomposing  $\ttt_a$,
$\ttt_b$ and $\Delta^\beta$ according to $\G=(\G_-+\G_0+\G_+)$, we first 
deduce that $\Delta^{\bt}$ must be $\G_+$-valued. Parametrizing
$$%\be
\Delta^{\bt} =- N^{\beta\rho} {\cal B}_{\rho\sigma} V^{\sigma} 
$$%\ee
we then find that (\ref{eq02}) is equivalent to ${\cal B}$ being 
anti-symmetric.
These solutions to the homogeneous equation for $\Delta^\beta$ correspond
to redefinitions of the $p^\beta$ of the form 
$$%\be
p^{\beta} \rightarrow p^{\beta} + N^{\beta\lambda} N^{\rho\mu} 
{\cal B}_{\mu\lambda} 
\dif q_{\rho}
$$%\ee
in the sense that such a redefinition can be  absorbed in 
the change  
$\Omega^\beta \rightarrow (\Omega^\beta + \Delta^\beta)$ in (\ref{qwak}).
This completes the analysis of the double pole. 
A further analysis of the
single pole shows that $I$ indeed satisfies (\ref{requi}), if
$N^{\beta\lambda} N^{\rho\mu} {\cal B}_{\mu\lambda}$ is  restricted to
be of the form $\dif^{\beta} A^{\rho} - \dif^{\rho} A^{\beta}$. 
Finally, using the identities in Appendix A, 
one verifies by an explicit calculation  that $\tr(I^2)$
is given by the expression in (\ref{sugawara}). 
%This completes  the proof of Theorem 3. 
{\em Q.E.D.}

\smallskip\noindent
{\em  Remark 3.}
The level shifts given in the theorem have already been found in the
second article in \cite{FF} (see also the appendix in \cite{FFR}).
The fact that $\Omega^\bt$ in (\ref{omega}) is $\G_-$-valued follows from
the upper triangularity of the coordinates.
We shall see in Section 5 that our formula for  $\Omega^\bt$  
reproduces all the earlier explicit results for the Wakimoto realization.

\smallskip\noindent
{\em  Remark 4.}
Let  $\rho_\G$ and $\rho_{\G_0^i}$ denote the Weyl vectors 
of the  Lie algebras $\G$ and $\G_0^i$ for $i>0$, respectively, 
and set $\rho_{\G_0}:=\sum_{i>0} \rho_{\G_0^i}$.
Identifying them as elements of the Cartan subalgebra,
the Weyl vectors by definition satisfy $\tr(\rho_\G H_{\alpha_l})=1$ 
for any simple root $\alpha_l$ of $\G$, 
and respectively $\tr(\rho_{\G_0^i} H_{\alpha_l})=1$ for any 
such simple root for which the Chevalley generator $H_{\alpha_l}$
lies in $\G_0^i$.
Since $\rho_\G$ is half the sum of the positive roots of $\G$,
the equality ${1\over 2}[V^\al, V_\al]= (\rho_{\G} - \rho_{\G_0})$ 
follows by evaluating the sum on the left hand side using  bases of
$\G_\pm$ that consist   of root vectors.
The defining properties of the Weyl vectors 
imply that $(\rho_\G-\rho_{\G_0})$ belongs 
to the abelian factor $\G_0^0\subset \G_0$ in (\ref{reductive}),
and thus  $\rho_\G=(\rho_\G-\rho_{\G_0})+\sum_{i>0} \rho_{\G_0^i}$ contains
pairwise orthogonal terms.
Taking this into acccount and applying the strange formula 
$24 \vert \rho_{\G}\vert^2 = \vert \psi \vert^2 h^* {\rm dim}(\G)$,
as well as its analogue for $\G_0^i$,
one readily checks that the central charges of the
constituent stress-energy tensors in (\ref{sugawara}) add up correctly
to ${k\,{\rm dim}(\G)}/({k+ h^*)}$.

\smallskip\noindent
{\em  Remark 5.}
The analysis of the uniqueness of  $\Omega^\beta(q)$ contained in the
proof above goes through in the same way with respect to  arbitrary 
polynomial coordinates on $G_-$.
We find that if a particular  $\Omega^\beta$ exists so that 
$I$ given by (\ref{qwak}) satisfies the current algebra, 
then the most general $\Omega^\beta$ having 
this property  arises from the replacement 
$$%\be
\Omega^\beta \rightarrow ( \Omega^\beta + \Delta^\beta)
\qquad\hbox{where}\qquad 
\Delta^\beta=  -N^{-1}_{\gamma\alpha} 
(\partial^\alpha A^\beta - \partial^\beta A^\alpha) V^\gamma 
$$%\ee
with   an arbitrary polynomial $A^\alpha(q)$.
This corresponds to a redefinition of the momenta  in (\ref{qwak}) 
similar to  (\ref{predef}).
These redefinitions are in fact canonical transformation since they
 preserve  the basic  OPEs.  Since  such canonical transformations
already exist at the classical level,  we see that there is 
no genuine ambiguity in the quantum correction  $\Omega^\beta$.

\section{Quantizations and  polymorphisms }
\setcounter{equation}{0}

In Section 3 we quantized the classical Wakimoto 
current (\ref{affwak}) in special polynomial coordinates on $G_-$, namely
in upper triangular ones.
It is natural to ask  whether the quantization can be performed 
in other coordinates as well, and whether the  quantizations
arising in different coordinates are essentially the same.
In this section we shall give a partial answer to these questions.
We call  two systems of global, holomorphic coordinates on $G_-$
{\em polynomially equivalent} if the change of coordinates is
given by a polymorphism. We  shall demonstrate that polymorphisms 
induce equivalent quantizations of the classical Wakimoto realization.

A  map $f: {\bf C}^n\rightarrow {\bf C}^n$ 
is a  {\em polynomial map} if its components are given by polynomials.
A  polynomial map is a {\em polymorphism} 
if it is one-to-one, onto and the  inverse is also a polynomial map.
It is known  that every injective polynomial map on ${\bf C}^n$ is a 
polymorphism \cite{BR,Ru}.  Let us also mention the very 
non-trivial\footnote{\ninerm  
It is obvious that  $\det(Df)$ is a  non-zero 
constant for any polymorphism $f$. 
The  converse is  open at present. }

\noindent
{\bf Jacobian conjecture \cite{Ke,BCW}.}
{\it A polynomial map $f:{\bf C}^n\rightarrow {\bf C}^n$ 
is a polymorphism  if and only if the Jacobian 
determinant $\det(Df)$ is a non-zero constant .}

Suppose  that $q_\alpha(Q)$  
describes the change of coordinates between 
two global, holomorphic systems of 
coordinates  $\{ q_\alpha\}$   and $ \{ Q_\alpha\} $ on $G_-$.
At the classical level,  one has  the  canonical transformation  
that maps $(Q, P)$ to $(q,p)$ according to 
\be\label{clcan}
q_\alpha = q_\alpha(Q),
\qquad
p^\alpha =p^\alpha(Q,P)={\dif Q_\mu \over \dif q_\alpha} P^\mu   , 
\ee
which follows geometrically if one thinks 
of $p_\alpha$ as  $-{\partial\over \partial q_\alpha}$.
It is  clear that the classical Wakimoto current in (\ref{affwak}) 
is a coordinate independent  object,  that is,  
$I(Q, P, j_0)= I(q(Q), p(Q,P), j_0)$.

Suppose now   that  the classical Wakimoto realization has been
quantized  in the $q_\alpha$-system in the sense that 
a polynomial $\Omega^\beta(q)$ was  found for which 
$I(q,p,j_0)$ in (\ref{qwak}) satisfies (\ref{requi}).
Below we explain how to construct a quantum field theoretic
--- vertex algebraic ---   analogue of the classical canonical 
transformation
(\ref{clcan}) 
in the case for which the change of coordinates is given by polynomials.
%%(at the formal level  we can do this for any change of coordinates).
After finding  this  `quantum polymorphism', we will simply 
define the quantized Wakimoto current in the $Q_\alpha$-system by 
\be\label{Qqwak}
I(Q,P,j_0)\equiv  I(q(Q), p(Q,P), j_0).
\ee
Then $I(Q,P,j_0)$ will have the form (\ref{qwak}) for some $\Omega^\beta(Q)$
induced by the construction, and 
hence this  quantization  is essentially unique due to the remark at the end 
of Section 3. 

We search for the quantum mechanical analogue of  (\ref{clcan})  in the form
\be\label{qcan}
q_\alpha=q_\alpha(Q),
\qquad
p^\alpha=p^\alpha(Q,P)=\left(P^\mu \Gamma(Q)^\alpha_\mu\right) + 
\Theta(Q)^{\al,\mu}\dif Q_\mu,
\ee
where classically $q_\alpha(Q)$ is a polymorphism.
This  formula takes into account the  ambiguity in normal ordering  
(\ref{clcan}) and it guarantees the invariance of the form of $I$ 
in (\ref{qwak}).
The standard OPEs in the $q_\alpha$-system (\ref{pqOPE}) 
must hold as a consequence of those in the $Q_\alpha$-system:
$$%\be
P^{\al}\underhook{(z) Q_{\bt}}(w) = \frac{-\delta^{\al}_{\bt} }{z-w}. 
\label{pqOPE2}
$$%\ee
Next we show that this condition determines  the functions 
$\Gamma(Q)^\al_\mu$ and $\Theta(Q)^{\al,\mu}$.

We first require that the OPE of $p^{\alpha}$ with $q_{\beta}$ has 
the correct
form. This determines $\Gamma$ as
$$%\be\label{Gamma}
\Gamma^\al_\mu=\frac{\dif Q_\mu}{\dif q_\al}.
$$%\ee
We then require the double pole of the OPE of $p^\al$ with $p^\beta$ 
to vanish, which admits the particular solution 
$\Theta^{\al,\mu}=\Theta^{\al,\mu}_*$ with 
$$%\be 
\Theta^{\al,\mu}_*=\frac{1}{2} 
\frac{\dif Q_{\rho} }{\dif q_{\al}} 
\frac{\dif^2 q_{\delta} }{\dif Q_{\lambda} \dif Q_{\rho} }
\frac{\dif^2 Q_{\lambda} }{\dif q_{\delta} \dif q_{\theta}} 
\frac{ \dif q_{\theta}}{\dif Q_\mu} .
$$%\ee
The general solution of the same requirement is found to be 
$$%\be
\Theta^{\al,\mu}= \Theta_*^{\al,\mu} + 
\frac{\dif Q_\eta}{\dif q_\al} B^{\eta, \mu},  
$$%\label{Theta}\ee
where $B=B^{\al,\beta}(Q)dQ_\al dQ_\bt$ is 
an arbitrary 2-form on $G_-$, $B^{\al,\bt}=-B^{\bt,\al}$.
The question is whether we can choose this 2-form in such a way 
that the simple pole in the OPE of $p^\al$ with $p^\beta$ vanishes. 
Introducing the matrix $C^{\al}$ by
$$%\be
(C^{\al})^{\eta}_{\rho}\equiv \frac{\dif Q_{\lambda}}{\dif q_{\rho}}
\frac{\dif}{\dif q_{\alpha}} \left( 
\frac{\dif q_{\eta}}{\dif Q_{\lambda} } \right) 
$$%\label{M}\ee
and using the preceding  formulae, we obtain   
$$%\be
p^{\al}\underhook{(z) p^{\bt}}(w)
= \frac{S^{\al,\beta}}{z-w},
$$%\label{pp}\ee
where 
$$%\be
S^{\al,\bt}=
\dif q_{\gamma} \left(\frac{1}{2} \tr( [C^{\al},C^{\bt}] C^{\gamma} ) 
- (dB)^{\al,\bt,\gamma} \right)
$$%\label{S}\ee
with
$$%\be
(dB)^{\al,\bt,\gamma}=
\frac{\dif Q_a}{\dif q_\al} \frac{\dif Q_b}{\dif q_\bt} 
\frac{\dif Q_c}{\dif q_\gamma} 
\left( \frac{\dif B^{a,b}}{\dif Q_c} 
+\frac{\dif B^{b,c}}{\dif Q_a} 
+\frac{\dif B^{c,a}}{\dif Q_b}\right).
$$%\ee
The notation is justified by the fact that 
$dB=(dB)^{\al,\bt,\gamma}dq_\al dq_\bt dq_\gamma$ for the 
2-form $B$ on $G_-$.
Since the singular part of the  OPE of $p^\alpha$ with $p^\beta$ must 
vanish, 
we conclude from the above that 
there exists a quantum canonical transformation from the 
$Q_\alpha$-system of coordinates to the $q_\alpha$-system if and only if  
$$%\be
E\equiv \tr([C^{\al},C^{\bt}]C^{\gamma})dq_\al dq_\bt dq_\gamma 
$$%\ee 
is an exact 3-form on $G_-$. 
In view of fact that $G_-$ has trivial topology,
this is equivalent to $E$ being closed. Now $E$ can be rewritten
as $E=\tr([\Gamma d\Gamma^{-1},\Gamma d\Gamma^{-1}],\Gamma d\Gamma ^{-1})$, 
a form which reminds
one of the topological term in  WZNW action, and when written in
this form one easily verifies that $dE=0$. Thus there always
exists a $B$ such that $dB=\frac{1}{2}E$.
It is also clear  that the components  of $B$ can be chosen to be 
polynomials,
although it may not be possible to write them down explicitly.
Choosing a  solution for $B$, 
the resulting canonical transformation is unique up
to a 1-form $A=A^\al(Q) dQ_\al$, which may be used to modify the 2-form $B$
appearing in the general solution for $\Theta^{\al,\mu}$   
according to $B\rightarrow B+dA$.
The meaning of this last ambiguity in $B$ is that it is possible
to redefine the momenta while keeping the coordinates fixed.
Such a canonical transformation is given by
$q_\al=Q_\al$, $p^\al = P^\al+(dA)^{\al,\bt}\dif Q_\bt$,
where $(dA)^{\al,\bt}= {\dif \over \dif Q_\al} A^\bt - 
{\dif \over \dif Q_\bt} A^\al$.
%$q_\al=Q_\al$, $p^\al = P^\al+B^{\al,\bt}\dif Q_\bt$,
%where the requirement is that $B$ is exact, 
%$B^{\al,\bt} = \dif^\al_Q  A^\bt - \dif^\bt_Q A^\al $
%with $\dif^\al_Q= {\dif \over \dif Q_\al}$.
These transformations, which  already made their appearance in 
Theorem 3 and in Remark 5,  are 
similarity transformations generated by the operator
$$%\be 
U= 
\exp\bigl( \oint {dz \over 2\pi i}\,A^{\al}(Q(z))\, 
\dif Q_{\al}(z)\bigr) .
$$%\ee

Having established the existence of the required 
 polynomials $\Gamma(Q)^\alpha_\mu$  and $\Theta(Q)^{\alpha,\mu}$,
we can implement the polynomial change of coordinates by the canonical 
transformation  (\ref{qcan}).
It is natural to enquire  
how the free field stress-energy tensor, $-p^{\al} \dif q_{\al}$,
behaves with respect to this  transformation.
A straightforward OPE calculation reveals  the transformation rule 
\be
-p^{\al} \dif q_{\al} = - P^{\al} \dif Q_{\al} + 
\frac{1}{2} \dif^2 \log\det \Gamma.
\label{extraT}\ee
The extra term  vanishes since $\det \Gamma$ is the Jacobian determinant 
of a polymorphism in our case.  
Incidentally, 
if the Jacobian conjecture is true,  then we could 
conclude from the same calculation that the form invariance 
of $p^\alpha \dif q_\alpha$ 
under a polynomial map $f$ requires $f$ to be a polymorphism.
It seems interesting that the free field form 
of the stress-energy tensor can be related to the polynomial 
equivalence of the underlying systems of coordinates in this way.
The  main result  of this section is now summarized as follows:

\medskip
\noindent{\bf Theorem 4.}
{\it  Let $q_\alpha$ and $Q_\alpha$ be 
polynomially equivalent systems of coordinates on $G_-$.
Suppose that the quantization of the classical Wakimoto realization
can be performed in the $q_\alpha$-system in such a way that
$I(q,p,j_0)$ has the form  (\ref{qwak}) with a
polynomial $\Omega^\beta(q)$, and the affine-Sugwara stress-energy tensor
 has the free field form given in Theorem 3.
Then, using the  quantum coordinate transformation (\ref{qcan}), 
$I(Q,P,j_0)$ in (\ref{Qqwak}) defines 
a quantization of the classical Wakimoto realization in 
the $Q_\alpha$-system  that has the same properties.}
\medskip

We  wish to conclude with some  remarks and conjectures 
related to the above result. 
Let us call a system of polynomial coordinates $q_\alpha$ on $G_-$ 
{\em admissible}  if  there exists a polynomial $\Omega^\beta(q)$ for which 
$I(q,p,j_0)$  in (\ref{qwak}) satisfies the current algebra.
Combining Theorem 4 and Remark 5, we see that the 
quantizations of the Wakimoto realization associated with 
two different systems of  admissible coordinates 
that are polynomially equivalent
can always be converted  into each other by a  canonical transformation
(\ref{qcan}).
Hence these quantizations  are essentially the same.
Theorem 3 tells us that all upper triangular
systems of coordinates, 
and therefore also all those  that are polynomially equivalent 
to upper triangular ones,  are admissible. 
It would be interesting to know 
whether all polynomial coordinates on $G_-$ are
polynomially equivalent to upper triangular ones or not.
We have the 

\smallskip
\noindent 
{\em Conjecture.} Any two upper triangular systems of coordinates are
polynomially equivalent. 
\smallskip

\noindent
{\em Remark 6.}
Let $\{ q_\alpha\}$ and $\{ Q_\alpha\}$ be two polynomially equivalent 
systems of coordinates on $G_-$ that are each upper triangular with 
respect to the same underlying gradation of the Lie algebra.
Let us further assume that $q_\alpha(Q)$ and $Q_\alpha(q)$ are 
{\em homogeneous} polynomials of degree $d_\alpha$ in their respective
arguments, which themselves carry the degrees specified in the notion 
of upper triangularity.
(We can verify this property in our examples, but have 
no general proof of it.)
Under this homogeneity property   
the object $\Theta_*$ vanishes  
and  $C^\alpha$ is an upper triangular 
matrix, which implies  that the 3-form $E$ vanishes too. 
Therefore the quantum polymorphism  
(\ref{qcan}) simplifies to 
$$
p^\al={\dif Q_\mu \over \dif q_\al} \left( P^\mu 
+ \left( \dif_Q^\mu A^\nu  - \dif_Q^\nu A^\mu\right) \dif Q_\nu \right),
\qquad
\dif_Q^\al = {\dif \over \dif Q_\al}. 
$$
In the present case 
this formula is free of normal ordering ambiguities.

\smallskip\noindent
{\em Remark 7.}
At the end of Section 2, we  explained the `composition 
property' of the classical Wakimoto realization.
Clearly, there exists a quantum mechanical version of this property.
To formulate a precise assertion, let us continue with the notations
in  (\ref{comp}) and parametrize $n_-$, $g_-$ and  $g_-^c$ by
upper triangular coordinates 
$\{ q^0_i\}$, $\{ q_\alpha\}$ and   $\{ Q_a\}$
 on  $G_{0,-}$, $G_-$ and on  $G_-^c$, respectively.
We can then quantize the classical Wakimoto realization
associated with the parabolic subalgebra ${\cal P}^c$ in two ways:
either directly in the $\{ Q_a\}$ system or by composing 
the quantized  Wakimoto realizations
associated with ${\cal P}_0$ and ${\cal P}$ that are 
obtained in the $\{ q^0_i\}$ and $\{ q_\alpha\}$ systems
of coordinates.
As a  consequence of the above,
the results of these two procedures are  related by a quantum 
canonical transformation whenever the $\{ Q_a \}$ system 
of coordinates on $G_-^c=G_{0,-}G_-$ is
polynomially equivalent to the system given by the union
$\{ q_i^0\} \cup \{ q_\alpha\}$. This is the case in all examples 
we know.
\smallskip

In conclusion, 
up to a few open problems like the polynomial 
equivalence of all upper triangular coordinates, 
we can argue from the results of this section that there is a 
unique quantization of the classical 
Wakimoto realization up to canonical transformations.
The fact that we could not completely prove the uniqueness 
of the quantization is due to the fact that in Theorem 3 
we required only a rather abstract condition on the coordinates,
whose content  is not easy to analyze completely. 
If we restrict the allowed coordinates to the 
polynomial equivalence class of the graded exponential 
coordinates, of the form in  
$g_-=\exp\left (\sum_\alpha q_\alpha E_{-\alpha}\right)$,
which are the ones used in practice, then 
our results already imply the expected uniqueness of the quantized 
Wakimoto realization.

%\newpage

\section{Examples of the Wakimoto realization} 
\setcounter{equation}{0}

We here show that the formulas of the Wakimoto current found 
in the literature previously follow as special cases of Theorem 3.
If needed in applications, one can easily work out other examples
along similar lines.

\subsection{Recovering the explicit formulas of Feigin and Frenkel}

In the first article of ref.~\cite{FF}, 
Feigin and Frenkel found the explicit 
form of the Chevalley generators $e_i(z)$, $h_i(z)$, $f_i(z)$  
for the principal Wakimoto realization of the $sl(n)$ current algebra. 
See also the second paper in \cite{BMP}. 
Next we reproduce the Feigin-Frenkel formula. 

In accordance with \cite{FF,BMP}, we parametrize the group element 
$g_-\in G_-$ by its matrix entries in the defining representation of 
$G=SL(n)$,
$$
g_-={\bf 1}_n + \sum_{n\geq a> b\geq 1} q_{a b} e_{a b }\,,
$$
where $e_{ab}$ is the usual elementary matrix.
Our task is to determine the projections of the current 
$I$ in (\ref{qwak})
on  the subspaces of $sl(n)$ with principal grades $0, \pm 1$.
Let us  decompose  $I$ as
$$
I={I}_{\rm geom} + Y
\qquad\hbox{with}\qquad 
I_{\rm geom}=-p^\beta N^{-1}_{\alpha \beta} g_-^{-1} V^\alpha g_-\,,
$$
where the index $\alpha$ is now a pair $\alpha=ab$ with $n \geq  a>b\geq 1$,
$V_{ab}=e_{ab}$, $V^{ab}=e_{ba}$, and 
$$
Y=g_-^{-1} j_0 g_- + K g_-^{-1} \partial g_- + 
g_-^{-1} \Omega^\beta g_- \partial q_\beta.
$$
Clearly, we have $Y_1=0$, $Y_0=j_0$. 
Specifying the formula (\ref{omega}) of $\Omega^\beta$ 
for the case at hand,
the grade $-1$ part of $Y$ turns out to be 
$$
Y_{-1}=\sum_{i=1}^{n-1} e_{i+1 i}\Bigl( (k+n-i-1)\partial q_{i+1 i}
 -q_{i+1 i} {\tr}(H_i j_0) \Bigr),
\qquad H_i\equiv e_{ii}-e_{i+1 i+1}.
$$

In order to determine $I_{\rm geom}$ we apply an indirect method
based on the finite group version of the Hamiltonian reduction.
The method utilizes the fact that $I_{\rm geom}$ is
the momentum map  for  the infinitesimal action of  $G$ on $T^* G_-$,
which is  induced from $G_-$ being the  big cell in the coset space
$P\backslash G$, where $P=G_+ G_0$ is the
 upper triangular subgroup of $G=SL(n)$.
We wish to determine the components of this  momentum map,   
$$
F^X(q,p)\equiv  {\tr}\left( X I_{\rm geom}\right)=
F^X_\alpha (q)p^\alpha 
\qquad 
\forall\, X\in sl(n).
$$
We see from the geometric picture of the Hamiltonian reduction
that the Hamiltonian vector field on $T^* G_-$ associated with $F^X$
is the lift of a corresponding  vector field, ${\cal V}(X)$, 
on $G_-$, which is given by  
$$
{\cal V}(X)=-F^X_{\alpha}(q) {\partial \over \partial q_\alpha}.
$$
${\cal V}(X)$ is obtained 
by  projecting   the  infinitesimal generator of the one parameter 
group action $(t, g) \mapsto g e^{tX}$on $G$
onto  the coset space $P\backslash G$,
and restricting the result to $G_-\subset P\backslash G$. 
In other words,  ${\cal V}(X)$ is the infinitesimal 
generator of the local one parameter group action  
$(t,g_-)\mapsto g_-^X(t)$ on $G_-$ defined 
by the Gauss decomposition
$$
g_- e^{tX}= g_+^X(t) g_0^X(t) g_-^X(t),\qquad
g^X_{0}(t)\in G_{0},\,\,\, g^X_{\pm}(t)\in G_{\pm}. 
$$
A  straightforward  computation 
(like the proof of  Theorem 2.4 in the second paper in \cite{BMP}) 
yields 
$$
{\cal V}(e_{l+1l})={\partial  \over \partial q_{l+1 l}} +
\sum_{i\geq l+2} q_{il+1} {\partial  \over \partial q_{il}}\,,
\qquad
{\cal V}(H)=\sum_{i>l} (\Delta_l - \Delta_i) q_{il} 
{\partial \over \partial q_{il}}
$$
for $H\equiv {\rm diag}(\Delta_1,\ldots, \Delta_n)$, 
%in particular %$$
%{\cal V}(H_l)= 2q_{l+1l} {\partial  \over \partial q_{l+1l}}
%+\sum_{i\geq l+2} \left( q_{il} {\partial \over \partial q_{il}}
%-q_{il+1} {\partial \over \partial q_{il+1}}\right)
%+\sum_{i\leq l-1} \left( q_{l+1i} {\partial \over \partial q_{l+1i}}
%-q_{li} {\partial \over \partial q_{li}}\right)\,,%$$
and
$$
{\cal V}(e_{ll+1})=\sum_{i\geq l+2} q_{il}{\partial \over \partial q_{il+1}}
-\sum_{i\leq l-1} q_{l+1i}{\partial \over \partial q_{li}} 
+q_{l+1l}\left( 
\sum_{i\leq  l-1} q_{li} {\partial \over \partial q_{li}}
-\sum_{i\leq l} q_{l+1i} {\partial \over \partial q_{l+1i}}\right).
$$
Collecting the above formulas, we arrive at the following 
expressions for the quantum fields  generating  the $sl(n)$ current algebra:
\bea
&&
{\tr}\left(e_{l+1l} I\right)
=-p_{l+1 l} -\sum_{i\geq l+2} p_{il} q_{il+1}
\nonumber\\
&&
{\tr }\left(H_l I\right)=
{\tr}(H_l j_0) -2 p_{l+1 l} q_{l+1 l} 
+\sum_{i\geq l+2} \left(p_{il+1} q_{il+1}  -p_{il}q_{il} \right)
+\sum_{i\leq l-1} \left( p_{li} q_{li} -p_{l+1i}q_{l+1i} \right)
\nonumber\\
&&{\tr}\left(e_{ll+1} I\right)=-{\tr}(H_l j_0)q_{l+1 l}
+(k+n-l-1)\partial q_{l+1l}
+\sum_{i\leq l-1} p_{li} q_{l+1i} - \sum_{i\geq l+2} p_{il+1} q_{il}
\nonumber \\
&& \qquad \qquad\qquad
+\sum_{i\leq l} p_{l+1i} q_{l+1i} q_{l+1 l} 
-\sum_{i\leq l-1} p_{li} q_{li}q_{l+1 l}\,.
\label{FF}
\eea
Here $p_{ab}$ denotes $p^\alpha$ for $\alpha=ab$ and normal ordering
is understood according to our conventions fixed previously.
If,  using  an automorphism of $sl(n)$, we  define
$$
e_i \equiv  {\tr}\left(e_{n-i+1\,,\, n-i} I\right),
\qquad
h_i \equiv - {\tr}\left(H_{ n-i} I\right),
\qquad
f_i \equiv  {\tr}\left(e_{n-i\,,\, n-i+1} I\right), 
$$
and  in addition set 
$$
\beta^{ab}\equiv  - p_{n+1-a\,,\, n+1-b}
\qquad\quad
\gamma^{ab}\equiv -q_{n+1-a\,,\, n+1-b}\,,
$$
and bosonize the current $j_0$ of the Cartan subalgebra 
of $sl(n)$ in the usual way, 
then  formula (\ref{FF})  becomes identical to the 
Feigin-Frenkel formula as described e.g.~in Theorem 3.2 
of the second article in \cite{BMP}.

For later reference, let us note  here the formula
\be\label{jforscr}
\tr\left(j e_{l+l l} \right) =  p_{l+1 l} + \sum_{1\leq  m < l}
(p_{l+1 m} q_{l m}), 
\ee 
which is obtained by using  that $\tr(j V_\al )$ corresponds to the
differential operator $-N^{-1}_{\al \bt} {\dif\over \dif q_\bt}$
whose action on $g_-$ generates the one parameter group 
$e^{-t V_\al} g_-$ for any $V_\al \in \G_-$.

\subsection{Recovering the formulas of Ito and Komata}

We now consider the principal case for an arbitrary Lie
algebra $\G$. We  here regard the vector fields on 
$G_-\subset P\backslash G$ as input data, and only search for 
the correction to these data for the currents 
associated with the Chevalley generators of $\G$.
The derivative term $K g_-^{-1} \dif g_-$ of $I$ in (\ref{qwak})
yields  an 
obvious correction. The only non-trivial quantum correction 
that will contribute is at grade $-1$, namely   
$$
\left(\Omega^\beta(q)\right)_{-1} \dif q_\beta. 
$$
As a preparation to determining this term, 
let us choose root vectors $E_{\pm \alpha}$ 
for every  positive root $\alpha \in \Phi^+$ 
in such a way that 
\be
[E_\al, E_{-\al}]= H_\al,
\qquad
[H_\al, E_{\pm \al}] = \pm 2 E_{\pm \al}.
\label{normal}
\ee
This implies (e.g.~\cite{H}) that 
$$
\tr\left(E_\alpha E_{-\beta}\right)={2\over \vert \alpha\vert^2} 
\delta_{\al\beta}, 
$$
and thus the dual bases of $\G_-$ and $\G_+$ can be  taken to be 
$$
V_\alpha= E_{-\alpha}\qquad \hbox{and}\qquad 
V^\alpha = {\vert \alpha\vert^2 \over 2} E_\alpha.
$$
For arbitrary roots $\gamma$, $\lambda$ for which $(\gamma+ \lambda)$
is a root, we have
$$
[E_{\gamma}, E_\lambda ] = C_{\gamma, \lambda} E_{\gamma +\lambda}
$$
with some non-zero $C_{\gamma, \lambda}$. 

We now parametrize $G_-$ by the exponential coordinates
$$
g_-= \exp\bigl(\sum_{\alpha\in \Phi^+}  q_\alpha E_{-\alpha}\bigr).
$$
In order to compute  the grade $-1$ part of 
$\Omega^\beta$ in (\ref{omega}), 
remember that the matrices  $N$ and $N^{-1}$ are block upper triangular, 
and their block-diagonal part is the identity. 
Thus  the  only contribution to $(\Omega^\beta)_{-1}$ will come from
the block-diagonal part of $N$ multiplied by  $\dif^\beta$
of the blocks of $N^{-1}$ just above the diagonal. Before taking the 
derivative,
these blocks of $N^{-1}$ are just the negative of the
 blocks of $N$ in the same position. 
If $A(x)$ is a matrix function, then 
$$
(\dif_x e^A) e^{-A} = \dif_x A + {1\over 2} [A, \dif_x A] + 
\hbox{higher commutators,}
$$
and we  see from this that
$$
{\dif^\alpha g_-} g_-^{-1}= N^{\alpha \lambda} E_{-\lambda} =
E_{-\alpha} + {1\over 2} \sum_{\gamma \in \Phi^+} q_{\gamma} 
[ E_{-\gamma}, E_{ -\alpha}] + 
\hbox{higher commutators.}
$$
Hence, for the blocks above the diagonal, we get  
$$
N^{\alpha \lambda}\vert_{ d_\lambda = d_\alpha +1} =
{1\over 2} q_{\lambda - \alpha} C_{\alpha - \lambda, -\alpha}.
$$
Therefore,
$$
\dif^\beta N^{-1}_{\alpha \lambda} \vert_{d_\lambda = d_\alpha +1}=
\delta_{\alpha + \beta, \lambda}
\left( -{1\over 2} C_{-\beta, -\alpha} \right).
$$
In particular, this is always zero if $\beta$ is not a simple root.
Now, 
$$
({\Omega^\beta})_{-1} = 
\dif^\beta N^{-1}_{\alpha \lambda}\vert_{d_\lambda = d_\alpha +1}\,
N^{\lambda\rho}\vert_{d_\lambda = d_\rho} [ V^\alpha, V_\rho]
$$
implies that $(\Omega^\beta)_{-1}=0$ unless $\beta$
is a simple root, and if $\beta$ is a simple root then 
$$
(\Omega^\beta)_{-1}= \sum_{\alpha, \alpha +\beta\in \Phi^+}
\dif^\beta N^{-1}_{\alpha, \alpha + \beta}\, N^{\alpha+\beta, 
\alpha +\beta}
[V^\alpha, V_{\alpha +\beta}].
$$
Substituting the above formulas, we get 
$$
(\Omega^\beta)_{-1}=E_{-\beta} 
\left(-{1\over 4} \sum_{\alpha, \alpha +\beta\in \Phi^+}
\vert \alpha \vert^2 C_{-\beta, -\alpha} C_{\alpha, -\beta -\alpha} 
\right).
$$
To summarize, for any  simple root $\beta$ we found  the quantum 
correction 
\be
\tr( E_\beta \Omega^\gamma \dif q_\gamma) = A_\beta \dif q_\beta
\quad\hbox{with}\quad 
A_\beta = -{1\over 2} 
\sum_{\alpha, \alpha +\beta\in \Phi^+}
{\vert \alpha \vert^2 \over \vert \beta\vert^2}
C_{-\beta, -\alpha} C_{\alpha, -\beta -\alpha}. 
\label{ITO}
\ee
Using that $\vert \alpha \vert^2
C_{\alpha, -\beta -\alpha} = \vert \beta\vert^2  
C_{-\beta -\alpha, \beta}$, which follows from (\ref{normal}),  
we can rewrite $A_\beta$ as 
\be
A_\beta =
{1\over 2} 
\sum_{\alpha, \alpha +\beta\in \Phi^+}
C_{-\alpha, -\beta} C_{-\beta-\alpha, \beta }.
\label{Iko}\ee
In fact, this  precisely is  
the quantum correction part of $\tr(E_\beta I)$ found by 
Ito and Komata \cite{IKo} (see also \cite{Ku}).
When making the comparison with eqs.~(24), (26) in \cite{IKo},
one should notice that the term in (26) there contains the  
contribution of $K g_-^{-1} \dif g_-$  too.
Note also that the normal ordering used in \cite{IKo}
is different from our convention, since 
there  $p$ is placed to the right and the normal ordering
is implemented  in a nested manner, like for example in $(q(qp))$.
Actually,   
the only term in the Chevalley generators where the 
normal ordering could be ambiguous is of this kind, and for this term
$
(q(qp))= (p (qq))
$.
We conclude that the formulas of \cite{IKo} are 
consistent with our result.

Finally, let us record the compact formula 
\be
A_\beta = {c_2(\G) \over 2 \vert \beta \vert^2} -1.
\label{Ito}\ee
This formula was found by Ito who verified it  by a case by case 
inspection 
of the Lie algebras using a Chevalley basis,
that is a Cartan-Weyl basis satisfying \cite{H,GOV}
$C_{\alpha,\beta}= C_{-\beta, -\alpha}$ in addition 
to (\ref{normal}),   see the second paper in \cite{IKo}.
In Appendix B  we present a simple derivation, 
which in particular shows that (\ref{Ito})  is valid in any Cartan-Weyl
basis subject to the  normalization condition (\ref{normal}). 

\smallskip\noindent
{\em Remark 8.}
In the very recent paper \cite{Ra}, 
explicit formulas  are given  for arbitrary components of the Wakimoto 
current in the principal case, using graded exponential coordinates.
Since the  above 
quantum correction for the Chevalley generators is recovered in  
\cite{Ra}, too, the formulas of \cite{Ra} must be consistent with
our result for $\Omega^\beta$ in (\ref{omega}),
which is valid with respect to an arbitrary
parabolic subgroup  $P\subset G$ and arbitrary upper triangular 
coordinates on $G_-$.

\subsection{A  simple example: $n=1+(n-1)$}

Next we consider a
partial free field realization  in which $\widehat{sl}(n)$ is
realized in terms of $\widehat{sl}(n-1)$, $\widehat{gl}(1)$ and
$(n-1)$ pairs of symplectic bosons.
This in principle allows for getting the  complete
free field  (principal Wakimoto) realization 
of $\widehat{sl}(n)$ by iterating  the construction (see also \cite{B}).

We choose the matrix $H\in sl(n)$ that determines the grading 
and the parabolic subalgebra to be
$$
H\equiv {1\over n}  {\rm diag}\left( (n-1), - {\bf 1}_{n-1}\right).
$$
Correspondingly, $\G_-$ is spanned by  the grade $-1$ elements
$e_{\alpha +1, 1}$, $\alpha=1,\ldots, (n-1)$, 
and is abelian.
Parametrizing  $g_-\in G_-$ as
$$
g_-= {\bf 1}_n + \sum_{\alpha =1}^{n-1} q_\alpha e_{1+\alpha, 1}\equiv  
{\bf 1}_n  + \left[\matrix{ 0&0\cr q & 0 \cr}\right],  
$$
we  find that
$
N_{\alpha \beta}= N^{-1}_{\alpha \beta}=\delta_{\alpha \beta}.
$
This implies  that the quantum correction 
$\Omega^\beta$ vanishes in this case, and that
$$
j(q,p)=\sum_\alpha p^\alpha  V^\alpha = 
\sum_{\alpha}  p^\alpha e_{1, 1+\alpha} \equiv
\left[\matrix{0&p \cr 0 & 0\cr}\right].
$$
The above notations mean that we think of $q$ and $p$ as 
$(n-1)$-component
column,  and row vectors, respectively, 
where the $n\times n$ matrices are written in 
an obvious block notation associated with the grading.
Using this notation  
the current $j_0\in \G_0$ can written in matrix form as
$$
j_0=  \xi nH + 
\left[\matrix{ 0&0 \cr 0 & \eta}\right],
$$
where $\eta$ belongs to $sl(n-1)$ and $\xi$ is a $gl(1)$ current.
We then parametrize the classical current $I\in sl(n)$  as
\be
I= g_-^{-1} ( - j + j_0) g_- + K g_-^{-1} \dif g_- \equiv  
\left[ \matrix{ a& b \cr c & d \cr}\right].
\label{Imatrix}
\ee
Straightforward matrix multiplication  gives 
\bea
&& a= (n-1) \xi - p q \nonumber\\
&& b= -p \nonumber\\
&& c= q p q + \eta q - nq \xi + K \dif q  \nonumber\\
&& d= \eta - \xi {\bf 1}_{n-1} + qp,
\label{1,n-1}
\eea
where $q$ and $p$ are again understood as column and row vectors.
For example, we have
$
(qpq)_\alpha = q_\alpha p^\beta q_\beta.
$
Quantization is performed  by normal ordering, 
pulling $p$ to the left.

Notice that if  $n=2$ then  $\eta = 0$ and (\ref{1,n-1})
looks formally identical to the formula of the 
original Wakimoto realization described  
at the beginning of  the introduction
(the current $j_0$ in the introduction
may be identified with $2\xi$ in terms of the present notation).
Applying the construction to $sl(n-1)$, and so on, one  can  
iteratively obtain  the  complete free field realization of
$I\in sl(n)$.
This  appears to be an effective way  to derive completely
explicit formulas  for  the free field realization. 
The resulting final formula will be equivalent to that obtained `in
one step' as was  noted before. 
A curious interpretation  of this result is  that Wakimoto's
original formula contains every information about the complete free field
realization of $\widehat{sl}(n)$ in the sense that one needs  to use
that formula only  in every step of the iterative procedure.
%This  procedure  could be further utilized 
%in applications of the Wakimoto realization. 

\subsection{The general two-blocks case: $n=r+s$}

Straightforwardly  generalizing  the  preceding example,
we define
$$
H=H_{r,s}\equiv {1\over n}{\rm diag}\left( s {\bf 1}_r, -r {\bf 1}_s\right).
$$
Then $g_-$, $j$ can be parametrized similarly as before, 
and $\Omega^\beta$ again vanishes since $\G_-$ is abelian.
Now  $q$ and $p$ are 
$s\times r$ and $r\times s$ matrices 
containing the conjugate pairs  in transposed positions, respectively.
We parametrize $j_0$ as
$$
j_0 = \xi n H_{r,s} + \left[\matrix{\eta_r & 0 \cr 0&\eta_s}\right]
$$
with $\eta_r \in sl(r)$ and $\eta_s\in sl(s)$.
The formula of $I$ in (\ref{Imatrix}) 
becomes 
\bea
&& a= \eta_r + s \xi {\bf 1}_r  - p q \nonumber\\
&& b= -p \nonumber \\
&& c= q p q + \eta_s q - q\eta_r - nq \xi + K \dif q \nonumber\\
&& d= \eta_s  -r \xi {\bf 1}_{s} + q p.
\label{r,s}
\eea
%%which reduces to (\ref{1,n-1}) for $r=1$, $s=(n-1)$.
Quantization is achieved by  normal ordering and  choosing
the levels of $\eta_r$, $\eta_s$ and $\xi$ according to 
Theorem 3.  
This example belongs to the series of  cases  
for which $P\backslash G$  is a hermitian 
symmetric space,  which has already been considered in \cite{B}.

\section{Screening Charges}
\setcounter{equation}{0}

In this section, we describe the screening charges relevant 
for the Wakimoto realization (\ref{qwak}).
The form of the screening charges has been known in the case 
where ${\cal G}_0$ is abelian \cite{BMP,FF,F1,Ku}.
 We are not aware of previous explicit results in the case where 
${\cal G}_0$ is nonabelian.

Screening charges are given by contour integrals of screening currents of 
conformal weight one. The screening currents, in turn, are constructed 
out of  polynomials in $p^{\al},q_{\al}$, and out of certain chiral 
primary fields for the affine Lie algebra generated by $j_0$. 
The crucial property of screening charges is that they commute with the 
Wakimoto currents given in (\ref{qwak});  more precisely, 
for generic $K$, the centralizer 
of the screening charges in the chiral algebra generated by 
$p^{\al},q_{\al}$ and $j_0$ should be generated precisely by the 
Wakimoto currents. 
Screening charges play an important role in the 
construction of irreducible representations and in the computation of 
correlation functions \cite{FF,BF,BMP,ATY,Ku}.
We will be able to define screening charges that commute with the 
Wakimoto currents for general ${\cal G}$ and ${\cal G}_0$. 
However, we can prove  \cite{F2} 
that the centralizer of these screening 
charges is generated by the Wakimoto currents only in the case where 
${\cal G}_0$ is abelian. 
For nonabelian ${\cal G}_0$, we conjecture that a similar statement 
holds. 
In the case of  nonabelian ${\cal G}_0$, a further problem
is that no explicit expression is known for the chiral primary fields  
of the current algebra generated by $j_0$, which  are contained in the 
screening operators.
It could  be useful  to  give a  free field 
realization of  these chiral vertex operators using a further Wakimoto 
realization for ${\cal G}_0$ with respect to the Cartan 
subalgebra of ${\cal G}_0$. 
%This might also lead to a construction of a vertex operator for the chiral 
%group-valued field that appears in WZNW theory, and reveal some of
%its `hidden' quantum group symmetry \cite{MR}. 
For abelian ${\cal G}_0$, one can bosonize ${\cal G}_0$ in terms of free
scalars and the chiral vertex operators are  given by exponents 
of these free scalars.

In the literature (see \cite{Ra} and references therein), 
one sometimes encounters formal screening operators 
containing negative powers of fields, especially in the analysis of 
admissible representations. 
Their precise definition involves further field redefinitions and
 bosonizations, and we will not consider them here.

We next describe the classical version of the screening charges, 
looking only at the property that they should commute with the Wakimoto 
currents. 
The required chiral primary fields can classically be obtained by taking a 
path-ordered exponent of the ${\cal G}_0$-valued current.
We then consider the quantum case, and discuss the 
centralizer of the screening charges. 
Finally, we briefly explain how all this is connected to 
free field realizations for $\W$-algebras.

\subsection{Screening charges: the classical case}

To define the classical screening charges, we need to introduce a 
non-periodic $G_0$-valued field $h_0(\sigma)$ 
which is related to $j_0$ by 
\be j_0 = K h_0^{-1} h_0' .
\label{defh}
\ee
It follows that $h_0$ satisfies $h_0(2\pi) = h_0(0)\cdot m$, where
$m$ is the monodromy of $j_0$ on the circle, 
$$%\label{defmm}
m=P\exp \left( \frac{1}{K} \int_0^{2\pi} j_0(\si) d\si \right). 
$$%\ee
We require $h_0$ to be a primary field with respect to $j_0$, 
\be\label{j0h0PB}
\{ \tr(Y_i j_0)(\sigma), h_0(\bar \sigma)\} =  h_0(\bar \sigma) Y_i\,
\delta(\sigma -\bar \sigma)\qquad
 Y_i\in \G_0,
\ee
 and to have vanishing PBs with $q_\al$, $p^\bt$. 
The affine current algebra PBs (\ref{j0PB})  of $j_0$ follow from 
(\ref{defh}) and (\ref{j0h0PB}). 
We can therefore consistently replace the variables $(j_0, q_\al, p^\bt)$ 
with the new variables $(h_0, q_\al, p^\bt)$.
Since the solutions of the differential 
equation (\ref{defh}) for $h_0$ are parametrized by the arbitrary initial 
value $h_0(0)$, the mapping $h_0 \mapsto j_0$ is many-to-one.
The PBs of $h_0$ with itself have a quadratic structure given in terms of 
a classical $r$-matrix, as described e.g.~in \cite{FL} and
references therein.  We will not need them explicitly.

We now show that the screening currents are some of the
components of $h_0 j h_0^{-1}$.
For any $\xi\in \G_-$,  define the current $\S_\xi$ 
and the associated charge $S_\xi$ by 
\be\label{clscr}
\S_\xi \equiv 
\tr(\xi h_0 j h_0^{-1})
\qquad\quad  
S_{\xi} \equiv \int_0^{2\pi} d\si\, {\cal S}_\xi(\sigma).
\ee

\smallskip
\noindent{\bf Theorem 5.}
{\it The charge $S_\xi$ in (\ref{clscr}) 
commutes with the classical Wakimoto current $I$ in 
(\ref{affwak}) if and only if $\xi \in \G_- \cap  [\G_+,\G_+]^{\perp}$.}

\medskip\noindent
{\it Proof.}
Let us compute the PB
\be
\{ \tr(T_a I ) (\si) , \tr(\xi h_0 j h_0^{-1} ) (\bar \si) \}.
\label{jj1}
\ee
The only unknown ingredient in this computation is the PB
of $I$ with $h_0$, which from (\ref{j0h0PB}) is found to be 
$$
\{  \tr(T_a I ) (\si), h_0(\bar \si) \} = (h_0 
\pi_0 ( g_- T_a g_-^{-1} ))(\si) \delta(\si -\bar \si).
$$
In addition, when working out (\ref{jj1}), one encounters $h_0'$,
which has to be replaced by $K^{-1} h_0 j_0$, using (\ref{defh}).
Putting everything together one obtains
\bea\label{screenPB} 
\{ \tr(T_a I ) (\si) , \tr (\xi h_0 j h_0^{-1} ) (\bar \si) \}
&=&  \tr( (h_0^{-1} \xi h_0) [j,\pi_+ (g_- T_a g_-^{-1} ) ] 
)(\si)\, \delta(\si -\bar\si)\nonu &&
+ K\,\tr( (h_0^{-1}\xi h_0) ( g_- T_a g_-^{-1}))(\si)\, 
\delta'(\si -\bar \si).
\eea
The first term on the right hand  side 
vanishes identically if and only if $\xi \in [\G_+,\G_+]^{\perp}$,
and since $\xi$ was an element of $\G_-$ to start with, this completes
the proof. {\it Q.E.D.}
\smallskip

Clearly, $\S_{\xi}$ in (\ref{clscr}) is a primary field with 
conformal weight one with respect to the stress-energy tensor 
(\ref{1.11}). 
For $\xi\in \G_- \cap  [\G_+,\G_+]^{\perp}$, we call 
$\S_\xi$ and $S_\xi$ the screening current and the   
screening charge associated with $\xi$.

It is important to notice that not all the screening currents are 
independent in general, since  
(\ref{clscr}) leads to the relation
\be\label{r0rel}
\S_\xi(r_0h_0, j) = \S_{r_0^{-1} \xi r_0} (h_0, j)
\qquad\quad\forall\,
r_0\in G_0.
\ee
The mapping $r_0: h_0 \mapsto r_0 h_0$ actually \cite{FL} 
defines a Poisson-Lie action of the group $G_0$ on the space of 
fields $h_0$, and (\ref{r0rel}) tells us that this action
transforms the screening currents $\S_\xi$ into each other 
according to the natural action of $G_0$ on the space 
$\G_- \cap  [\G_+,\G_+]^{\perp}$.
Hence, there exists  one independent screening charge for each
irreducible representation of $G_0$ in $\G_- \cap  [\G_+,\G_+]^{\perp}$.
Using the gradation in (\ref{grading}),
which satisfies 
$[H, E_{\alpha_l}]=n_l E_{\alpha_l}$ with $n_l\in \{0,1\}$
for any simple root $\alpha_l$ of $\G$, 
it is not difficult to  show that 
$$
\G_- \cap  [\G_+,\G_+]^{\perp}= \G_{-1}.
$$
The highest weight vectors of the irreducible representations
of $G_0$ in $\G_{-1}$ are those  root vectors 
of simple roots that lie in $\G_{-1}$, i.e., the 
$E_{-\alpha_l}$  for which $n_l=1$. 
Thus a set of independent screening currents can be chosen to contain 
$$
\S_{(l)} \equiv  \S_{\xi} \qquad \hbox{for} \qquad \xi = 
E_{-\alpha_l}\in \G_{-1}.
$$

In the next section, we shall quantize the classical 
screening currents $\S_{(l)}$.
For this, it will be advantegeous to rewrite them in the form 
\be\label{Sl}
\S_{(l)} = \tr( j M_{(l)})= j_\alpha M^\alpha_{(l)} 
\ee
with
\be\label{Ml}
M_{(l)}\equiv  h_0^{-1} E_{-\alpha_l} h_0. 
\ee
Here $j_\alpha = \tr(V_\alpha j)$, 
$M_{(l)}^\alpha = \tr(V^\al M_{(l)})$
with respect to dual bases 
$V_\alpha$ of 
 $\G_{-1}$ and 
$V^\alpha$ of $\G_{1}$.
The field $M_{(l)}(\sigma)$ takes its values
in the irreducible representation of $G_0$
built on the highest weight vector $E_{-\alpha_l}$,
which we denote by $\V_{(l)} \subset \G_{-1}$,
i.e., $M_{(l)}^\alpha$ vanishes if $V_\alpha \in \V_{(l')}$ for 
$l'\neq l$.
Incidentally, the $\V_{(l)}$  are pairwise inequivalent 
representations. 
The crucial property of $M_{(l)}$ is that, 
as  a consequence of (\ref{j0h0PB}), it  is  a primary field 
with respect to $j_0$, 
\be\label{clMlpr}
\{ \tr(Y_i j_0)(\sigma), M_{(l)}(\bar \sigma)\} = 
-[Y_i , M_{(l)}(\bar \sigma)] 
\delta(\sigma -\bar \sigma)
\qquad\quad  Y_i \in \G_0.
\ee

\subsection{Screening charges: the quantum case}

We now turn to the quantum screening charges. 
This is a more difficult case, because we cannot define an operator 
$h_0$ by means of (\ref{defh}).
However, according to (\ref{Sl})  we do not
need $h_0$ itself to construct the screening currents  $\S_{(l)}$,
only the chiral primary fields $M_{(l)}$.
The general theory of vertex operators (see e.g.~\cite{VA,TK})
associates with  each irreducible representation of $\G_0$ a unique 
vertex 
operator that creates the corresponding representation of the affine 
algebra (\ref{j0i}) of $j_0$ from the vacuum, and this is precisely 
sufficient for our purposes. 
We denote the chiral vertex operator associated with the
representation $\V_{(l)}\subset \G_{-1}$ by $M_{(l)}$, 
in components $M_{(l)}(z)=V_\al M_{(l)}^\al(z)$, 
to indicate that it is a quantization of the corresponding classical 
object.
It satisfies  the OPE that is the quantum version of the Poisson 
bracket (\ref{clMlpr})
\be\label{j0MlOPE}
\tr(Y_i \underhook{j_0)(z) M_{(l)} } (w) = 
\frac{-[Y_i,M_{(l)}(w) ]}{z-w}
\qquad\quad  Y_i \in \G_0.
\ee
In the case where $\G_0$ is abelian, 
the components of $j_0$ and the $M^{\al}_{(l)}$ can be represented 
by derivatives  and  exponentials of free scalar fields, respectively.

Using the chiral primary fields 
$M^{\alpha}_{(l)}$, we find that 
the screening currents in the quantum theory 
are given by the same equation as the classical ones,
namely by eq.~(\ref{Sl}) if we normal order as before by moving the 
momenta contained in $j_\al$ to the left.
In the rest of this section, we use upper triangular 
coordinates on $G_-$ as in Section 3, and adopt the definition
\be \label{qscr}
\S_{(l)}\equiv 
(p^{\bt} (N^{-1}_{\al\bt} M^{\al}_{(l)} ))
\qquad \quad \qquad 
S_{(l)} \equiv \oint \frac{dz}{2\pi i} \S_{(l)}(z).
\ee
Then the  following result holds.

\smallskip
\noindent{\bf Theorem 6.}
{\it
The OPEs between 
the screening current $\S_{(l)}$ in (\ref{qscr}) and the Wakimoto current
$I$ in (\ref{qwak}) have the form 
\be\label{screenOPE}
 \S_{(l)} \underhook { (z) \tr(T_a I) } (w) 
 = \frac{ -y\, \tr(g_- T_a g_-^{-1} M_{(l)} ) (w)}{(z-w)^2}
 \ee
where  $y=\frac{1}{2} |\psi|^2 (k+h^*)$ and $T_a$ is a basis of $\G$.
As a consequence, the screening charge $S_{(l)}$ 
commutes with the Wakimoto current.
Furthermore, the screening current $\S_{(l)}$ is a primary field
with conformal weight one with respect to the 
stress-energy tensor in (\ref{sugawara}).}

\medskip\noindent
{\it Proof.}
The verification of the OPE in (\ref{screenOPE}) 
is a long and tedious exercise, using the various
properties of $N^{-1}_{\al\bt}$, and the identities given in Appendix~A.
The only novel ingredient is that in the same way as we encountered 
$h_0'$ in the proof of Theorem 5, 
where we could replace it by $K^{-1} h_0 j_0$, 
we now encounter the operator $\dif M^{\al}_{(l)}$.
Chiral primary fields generating highest weight 
representations of affine algebras have a generic null vector, first 
found by Knizhnik and Zamolodchikov \cite{KZ}. This null vector
translates into an operator identity, which for $M^{\al}_{(l)}$ reads as
\be \label{jj3}
y \dif M^{\al}_{(l)}= 
\left((\tr [j_0,V^{\al}] V_{\mu}) M^{\mu}_{(l)}\right).
\ee
As for the conformal weight of $\S_{(l)}$, this is another
standard OPE calculation. 
It turns out that %in upper triangular coordinates 
the conformal weight of
$(p^{\beta} N^{-1}_{\alpha\beta})$ is one and  
the conformal weight of $M^{\al}_{(l)}$ is zero, 
completing the proof of the theorem. 
{\it Q.E.D.}
\smallskip
 
Observe that the OPE in (\ref{screenOPE}) 
corresponds precisely to the PB in (\ref{screenPB}).
It vanishes for $T_a\in \G_{\leq 0}$, and
for  $T_a\in \G_{1}$ the conjugation by $g_-$ drops out.
In the principal case,  (\ref{screenOPE}) reproduces the known OPEs
between the screening currents and the  components of $I$ 
associated with the Chevalley generators of $\G$ \cite{BMP,Ku}.

Of course, 
in the principal case  the formula of the screening currents in
(\ref{qscr}) itself reduces  to  the  known result \cite{BMP,FF,F1,Ku}.
As an illustration, let us present  
the  detailed form  of the $\S_{(l)}$ for the example 
of the $sl(n)$ current algebra discussed in Section~5.1. 
In this case,
the abelian current algebra generated by $j_0$ can be bosonized by 
$j_0=-i \sqrt{y} \partial \phi$, where $\phi$ is a $\G_0$-valued scalar
field with the OPE
$$
\tr (Y_l \underhook{\phi)(z) \tr(Y_m} \phi)(w)  =
-\tr( Y_l Y_m) \log(z-w)
\qquad Y_l, Y_m\in \G_0, 
$$
and $\G_{-1}$ decomposes into the one-dimensional 
representations of $\G_0$ spanned by the root vectors
$E_{-\alpha_l}= e_{l+1 l}$ for $l=1,\ldots, n-1$.
The chiral vertex operators  $M_{(l)}$ are now 
given by $M_{(l)}= e_{l+1 l}
:\exp(-\frac{i}{\sqrt{y}} \alpha_l(\phi)): $.
Combining this with (\ref{Sl}) and (\ref{jforscr}),  we 
find the screening currents  
$$
\S_{(l)} = \bigl( p_{l+1l} + \sum_{1\leq m <l}
p_{l+1 m} q_{l m} \bigr) 
:\exp(-\frac{i}{\sqrt{y}} \alpha_l(\phi)): 
$$
consistently  with their  formula in \cite{BMP}.

In the case where ${\cal G}_0$ is abelian, one can define a so-called 
Wakimoto module $W_{\lambda}$ as being the Verma module generated 
by $j_0,p^{\alpha},q_{\alpha}$ from a highest weight state whose 
$j_0$-eigenvalue is the weight $\lambda$  \cite{Wa,FF}. 
The Wakimoto modules fit into a complex whose $m$th term $C^m$ 
is $C^m=\oplus_{s\in W,l(s)=m} W_{s(\rho)-\rho}$, where $W$ is the Weyl 
group of $\G$, $l(s)$ is the length of the Weyl group element $s\in W$, 
and $\rho$ is half the  sum of the positive roots. 
The differential of this complex is given by suitable multiple contour 
integrals of products of the screening currents. 
In particular, the differential $d_0$ mapping $C^0$ to $C^1$ is
given by the direct sum of the screening charges themselves. 
In \cite{F2} it is shown that the zeroth cohomology of this complex 
is for generic $K$ equal to the vacuum module of $\hat{\cal G}_K$, 
and all other cohomologies vanish. 
This means that the kernel of $d_0$ is the vacuum module of 
$\hat{\cal G}_K$, or in other words, the centralizer of the screening 
charges acting on the algebra generated by  $j_0,p^{\alpha},q_{\alpha}$ 
is precisely generated by the Wakimoto current $I$. 
Unfortunately, we do not know of a similar finite
resolution for nonabelian ${\cal G}_0$, but we conjecture that such a 
resolution should also exist, where the first differential is again given
by the direct sum of the screening charges. 
Such a resolution would immediately imply that 
for generic $K$ the Wakimoto current 
generates the centralizer of the screening charges $S_{(l)}$ 
in the nonabelian case as well,  but we have no rigorous proof of this 
statement at present.  

\subsection{Applications to $\W$-algebras}

With the explicit results for the Wakimoto realization and 
its screening charges, it is interesting to see what we can
learn from this for $\W$-algebras, since it has for example
been observed that application 
of a Hamiltonian reduction to the
Wakimoto realization leads to free field realizations and
resolutions for representations of $\W$-algebras \cite{BO,FF2,F1,F2,FO}.
Alternatively, free field realizations for $\W$-algebras can
be obtained by expressing the generators of the $\W$-algebra in
terms of those of an affine Lie algebra through the Miura map,
and by subsequently using a Wakimoto realization for this 
affine Lie algebra \cite{BT}. 
Quite remarkably, the level shifts found in the Miura map in \cite{BT} 
match precisely those appearing in Theorem~3. 
This is no coincidence, and we will now briefly explain the relation. 
%%between these various statements.

Consider those $\W$-algebras that can be obtained by  Hamiltonian 
reduction of an affine Lie algebra based on an $sl_2$ embedding, 
with a set of first class constraints that constrain the components of the
current that lie in $\G_+$ and generate the gauge group 
${\widetilde G}_-$ 
at the classical level. 
The corresponding quantum $\W$-algebra is by definition the BRST 
cohomology of 
\be \label{defq}
Q= \oint \frac{dz}{2\pi i} [( I_{\al} - \chi(I_{\al}) ) c^{\al} 
 - \frac{1}{2} f_{\al\bt}{}^{\gamma} (b_{\gamma}(c^{\al} c^{\bt} )) ]
\ee
acting on the chiral algebra generated by the current 
$I$ of $\widehat{\G}_K$ together with the ghosts and antighosts 
$c^{\al}$ and $b_{\beta}$.
Here $I_{\alpha} = \chi(I_{\alpha})$ are the first class constraints, 
$I_\alpha=\tr(V_\al I)$ for a basis $V_\al$ of $\G_-$
using the decomposition $\G=\G_- + \G_0 + \G_+ $ associated with 
the $sl_2$ embedding.  
For notation and the computation of this cohomology, see \cite{BT}. 
Let us now repeat the calculation of the cohomology, but first insert 
the Wakimoto form (\ref{qwak}) for $I$ in (\ref{defq}). 
In principle we could choose the underlying parabolic subalgebra 
arbitrarily, but the form of the BRST operator suggests that 
we should identify  the subalgebra 
$\G_+\subset \G$ that appears in the first class constraints in 
(\ref{defq})
 with the $\G_+$ that is used in the Wakimoto realization. 
This will indeed lead to a major simplification, and we restrict our
attention to this case.

Thus we are going to compute the BRST cohomology of $Q$ in (\ref{defq})
acting on the algebra generated by $j_0,p^{\al},q_{\al},b_{\al},c^{\al}$. 
Clearly, since $j_0$ does not appear at all in $Q$, it will
survive in the cohomology. In addition, we claim that the
cohomology on the other fields is trivial, so that the full
cohomology is in fact generated by $j_0$. To prove this, we note
that the algebra generated by $p^{\al},q_{\al},b_{\al},c^{\al}$ is the
same as the algebra generated by $\hat{p}_{\al},q_{\al},\hat{c}_{\al},
b_{\al}$, where $\hat{p}_{\al} \equiv Q'(b_{\al})$, with $Q'$ equal
to $Q$ without the $\chi(I_{\al})$ term, 
and
$$
\hat{c}_{\al}   = c^{\gamma} \tr(V_{\gamma} N^{-1}_{\bt\al} g_-^{-1}
V^{\bt} g_-) .
$$
That the transormation between these two sets of variables is
one-to-one and invertible follows from the  fact that the coordinates 
$q_{\al}$ are upper triangular. 
In terms of the new variables, the BRST transformations are
particularly simple,
$$
0 \rightarrow b_{\al} \rightarrow \hat{p}_{\al} - \chi(I_{\al}) 
\rightarrow 0
\qquad\qquad 
0 \rightarrow q_{\al} \rightarrow \hat{c}_{\al}  \rightarrow 0
$$
from which one immediately derives (`quartet confinement')
that the BRST cohomology is trivial \cite{BT}.

Clearly, the above BRST cohomology has not yet anything to do with
the $\W$-algebra, since it is generated by $j_0$. 
The $\W$-algebra is defined to be the BRST cohomology of $Q$ acting on  
a chiral algebra generated by the current algebra $\widehat{\G}_K$, 
and we have extended this current algebra to the algebra
of $j_0,q_{\al},p^{\al}$, explaining the apparent discrepancy.
Suppose now that   $\widehat{\G}_K$ 
is recovered as the centralizer of the screening charges $S_{(l)}$ in
(\ref{qscr}) acting on the algebra of $j_0,q_{\al},p^{\al}$.
Then we still have to take this into account to obtain the $\W$-algebra. 
The screening charges 
commute with the Wakimoto current and therefore should give
a well-defined action on the BRST cohomology. If we choose
as representatives for the cohomology anything generated by $j_0$,
then the action of the screening charges in (\ref{qscr}) will not
preserve this set of representatives. 
In order to obtain such screening charges that act inside this set of
representatives, we have to add BRST-exact pieces to the
screening charges in (\ref{qscr}).
Let us therefore take a closer look at the BRST cohomology.
If we define $r_{\alpha}=\hat{p}_{\alpha}-\chi(I_{\alpha})$,
then the action of $Q$ can be summarized as
$0\rightarrow b_{\alpha} \rightarrow r_{\alpha} \rightarrow 0$ and
$0\rightarrow q_{\alpha} \rightarrow \hat{c}_{\alpha} \rightarrow 0$.
We used this to argue that the cohomology on the space generated by
$b,r,q,\hat{c}$ was trivial. In other words, every $Q$-closed 
differential polynomial
of $b,r,q,\hat{c}$ is in $Q$-cohomology equal to a constant. 
We now associate 
a weight to $b,r,q,\hat{c}$, namely we give $b$ and $r$ weight one, $q$ 
and $\hat{c}$ weight
zero and $\partial$ weight one. 
This then associates a weight to any differential polynomial expression 
in $b,r,q,\hat{c}$, but in the remainder we only look at polynomials 
whose weight is at most one.
Since the  BRST operator preserves the weight and the constants 
have weight zero, a differential
polynomial of weight one that is $Q$-closed is equal to zero in 
$Q$-cohomology. Any polynomial
of weight zero is an ordinary polynomial in the commuting variables $q$ 
and $\hat{c}$, and
restricted to these $Q$ can be identified with the usual exterior 
derivative on a $\dim(G_-)$-dimensional
plane. In that case the Poincar\'e lemma tells us that every closed 
form is exact except for the
constants. Altogether this shows that any polynomial $P(b,r,q,\hat{c})$ 
of weight at most
one is in $Q$-cohomology equal to $P(0,0,0,0)$. 

Returning to the screening charges $S_{(l)}$ in (\ref{qscr}), 
it is easy to see that they can be written as
$$
S_{(l)} = \oint \frac{dz}{2\pi i} 
(p^{\bt} (N^{-1}_{\al\bt} M^{\al}_{(l)} )
 ) = \oint \frac{dz}{2\pi i} 
((\chi(I_{\alpha}) + f_{\alpha}(b,r,q,\hat{c}) )M^{\al}_{(l)} ),
$$
where $f_{\alpha}$ is some polynomial of weight one.
Hence  we can apply
the result derived in the previous paragraph and deduce that $S_{(l)}$ is 
in $Q$-cohomology the same as 
$$
S^{\W}_{(l)} \equiv \oint \frac{dz}{2\pi i} 
(\chi(I_{\alpha}) M^{\al}_{(l)} ),
$$
and these operators do act in the space generated by $j_0$. 

However, the above result does not yet imply that for generic $K$
 the centralizer of the screening charges $S^{\W}_{(l)}$ acting on
the algebra generated by $j_0$ is indeed the $\W$-algebra. 
To prove such a statement, we would need a resolution of the vacuum 
module of $\hat{\cal G}_K$ as mentioned in the preceding subsection. 
With such a resolution, the $\W$-algebra would by definition be 
$H^{\ast}_Q(H^{\ast}_d(C^{\ast} \otimes {\cal A}_{\rm ghosts}))$, with 
$d$ the differential of the resolution and ${\cal A}_{\rm ghosts}$ the 
algebra generated by $b$ and $c$.
This is the second term in a spectral sequence calculation of 
$H^{\ast}_{d+Q}$ which degenerates at the second term. 
We can then compute $H^{\ast}_{d+Q}$ using the opposite spectral sequence
with second term $H^{\ast}_d (H^{\ast}_Q)$. 
This spectral sequence must also degenerate at the second term, and from 
this point of view the cohomology is equal to the centralizer of the
$S^{\W}_{(l)}$ acting on the algebra generated by $j_0$. 
For abelian ${\cal G}_0$ the resolution exists \cite{F2} 
and this argument proves 
that the $\W$-algebra is the centralizer of the $S^\W_{(l)}$ acting on the 
algebra generated by $j_0$,
and everything can easily be represented in  terms of free fields.
For nonabelian ${\cal G}_0$ a similar statement follows 
once we assume the 
existence of the conjectured resolution.

\newpage

\section{Conclusions}

In this paper we have provided explicit expressions for general Wakimoto 
realizations of current algebras. 
The advantages of our approach are that (i) we obtain the form of   
all components of the $\G$-valued current, not just those corresponding 
to the Chevalley generators of $\G$, 
(ii) it has a clear geometrical origin and 
(iii)  we obtain the realizations for nonabelian 
${\cal G}_0$ at one stroke as well. 
Moreover, our explicit formulas are  valid 
in arbitrary upper triangular coordinates, and we have 
determined the screening charges for all cases, too. 
Wakimoto realizations play an important role in the representation 
theory of affine Lie algebras 
as the building blocks
of resolutions of irreducible highest weight representations 
\cite{FF,BF,BMP}, where the screening currents 
are used to construct intertwiners between different Wakimoto modules. 
Furthermore,
Wakimoto realizations can be used to compute correlation functions in the 
WZNW model \cite{FF,BF,BMP,ATY,Ku},
and to obtain free field realizations of ${\cal W}$-algebras 
\cite{BO,FF2,F1,F2,FO}. 
For each of these applications, our
explicit expressions should be useful if one wants to derive general 
results (as we illustrated
for ${\cal W}$-algebras above),  or if one simply wants to work out 
a complicated example in detail. 
Let us also mention integrable deformations of conformal field theory,
Toda theories  and generalized KdV hierarchies  
as subjects where our results might prove useful.

In addition to  the above-mentioned
applications, there are various issues in the construction itself that we 
would like to understand better. 
One of them is the role the screening charges play in
the Hamiltonian reduction. 
In this paper we introduced them in a somewhat ad hoc manner.
A more conceptual derivation of the screening charges 
should probably make use of a left-right separated version of the WZNW 
phase space and its Poisson-Lie geometry. 
Another very interesting problem is whether we can obtain  an explicit 
expression for the 
chiral group-valued field of WZNW theory in terms of free fields. 
Classically, there are
several chiral group-valued fields that one can consider, 
distinguished by the different
monodromies  they possess, with different 
Poisson brackets. Upon quantization,
quantum group structures may arise, and we would like to know what kind 
of quantum
group structure, if any, will naturally appear in a Wakimoto-type free 
field realization of
the chiral group-valued field. These may be related to the quantum group 
structures encountered
in \cite{BMP}, and we hope to come back to this point in the future.

\bigskip
\bigskip

\noindent
{\large \bf  Acknowledgements.}
LF is  grateful to B.~Feigin, E.~Frenkel, K.~Ito and G.~Meisters 
for useful conversations and correspondence. 
He also wishes to thank I.~Tsutsui for hospitality at the Institute
for Nuclear Studies, University of Tokyo during the course of this work.
JdB is supported in part by the Director, Office of Energy
Research, Office of Basic Energy Services, of the US Department
of Energy under Contract DE-AC03-76SF00098, in part by the
National Science Foundation under grant PHY95-14797, and
is a fellow of the Miller Institute for Basic Research in Science.

\newpage

\appendix

\setcounter{equation}{0}

\renewcommand{\theequation}{A.\arabic{equation}}

\section{Some OPE identities}

In this appendix we record  some  identities for OPEs of composite fields
formed out of $p^\alpha$, $q_\beta$ that are subject to the OPEs
in (\ref{pqOPE}).
These identities were used in our computations. 
They follow straightforwardly from the general OPE
rules given in the appendix of \cite{BBSS}.
Recall that the normal ordered product $(AB)$ of two fields $A$ and $B$
is defined  by
$$
\left(AB\right)(w) = {1\over 2\pi i} \oint { dz \over z-w}A(z) B(w)
$$
using a contour that winds around $w$ counterclockwise.
Then for polynomial functions $f$ and $g$ of the $q_{\bt}$, but not of
their  derivatives, the Wick rule implies 
\begin{eqnarray}
p^{\al}(\underhook{z) f(}w) & = & \frac{-\dif^{\al} f(w)}{z-w} \nonu
f(\underhook{z) (p^{\al}g)(}w) & = & \frac{(g\dif^{\al}f)(w)}{z-w} \nonu
p^{\bt}(\underhook{z) (p^{\al}g)(}w) & = 
& \frac{-(p^{\al}(\dif^{\bt}g))(w)}{z-w} \nonu
(p^{\al}f)(\underhook{z) (p^{\bt}g)(}w) & = & 
\frac{-(\dif^{\bt}f(z) \dif^{\al}g(w))}{(z-w)^2} + 
\frac{(p^{\al}(g \dif^{\bt} f)-p^{\bt} (f \dif^{\al} g))(w)}{z-w} \nonu
% \\[.5mm]  \nonu
(p^{\al}f)(\underhook{z) (g\dif q_{\bt})(}w) & = & 
\frac{-(f(z)g(w))\delta^{\al}_{\bt} }{(z-w)^2} +
\frac{-(f\dif^{\al}g \dif q_{\bt})(w)}{z-w}.
\label{A.1}
\end{eqnarray}
Here a  classical object $p f(q)$ is represented by the normal ordered
object $(p(f(q)))$,  which we  denote  as  $(p f)$ for simplicity,
but having this particular ordering in mind.
As in the main text, 
$\dif^\alpha  = {\dif \over \dif q_\alpha}$
and  $\dif$ means derivation  with respect to the complex parameter, 
$(\dif f)(z)={\dif f\over\dif z}$.
To verify the form of the affine-Sugawara stress-energy tensor in 
Theorem 3,  we  have further 
used the following rearrangement identities for normal ordered products:
\begin{eqnarray}
\Bigl((p^{\al} f)( p^{\bt} g)\Bigr) & = & 
-\frac{1}{2} (\partial^{\al} g) \dif^2 (\dif^{\bt} f )
 +\Bigl( \dif p^{\al} (g \dif^{\bt} f)\Bigr) + 
\Bigl(p^{\al} (g \dif \dif^{\bt} f) \Bigr)
\nonu & & -\Bigl(p^{\bt}( \dif f \dif^{\al} g)\Bigr)
 +\Bigl( p^{\bt} (p^{\al} (gf))\Bigr) \nonu
\Bigl((p^{\al} f)g\Bigr) & = & -\dif f \dif^{\al} g + 
\Bigl(p^{\al}(gf)\Bigr) \nonu
\Bigl(g(p^{\al} f)\Bigr) & = & f \dif \dif^{\al} g + 
\Bigl(p^{\al} (gf)\Bigr) \nonu
\Bigl((p^{\al} f)(g \dif q_{\bt})\Bigr) & = & 
-\dif f (\dif^{\al} g) \dif q_{\bt} 
-\frac{1}{2} \delta^{\al}_{\bt} g \dif^2 f + 
\Bigl(p^{\al}(fg\dif q_{\bt})\Bigr) \nonu
\Bigl((g \dif q_{\bt}) (p^{\al} f)\Bigr) & = & 
 f \dif((\dif^{\al} g) \dif q_{\bt} )
-\frac{1}{2} \delta^{\al}_{\bt} f \dif^2 g + 
\Bigl(p^{\al}(fg\dif q_{\bt})\Bigr). 
\label{A.2}
\end{eqnarray}

\newpage

\setcounter{equation}{0}

\renewcommand{\theequation}{B.\arabic{equation}}

\section{Derivation of Ito's formula for  $A_\beta$}

The purpose of this appendix is to derive the formula of 
$A_\beta$ in (\ref{Ito}) from that in (\ref{Iko}).  
First,  let us introduce the notation 
$E^\alpha \equiv {\vert \alpha \vert^2\over 2} E_{-\alpha}$
for any root $\alpha$. 
We use the normalization given in (\ref{normal}), and hence 
$E^{\alpha}$ is the dual of $E_{\alpha}$ with respect to the inner 
product $\tr$ of $\G$, that is 
$\tr(E_{\alpha} E^{\bt}) = 1$ iff  $\alpha=\bt$ 
where $\alpha,\bt$ are arbitrary elements of the root system $\Phi$. 
We also introduce a  basis $\{H_i\}$ for the Cartan subalgebra,
with dual basis $\{H^i\}$ with respect to $\tr$, $\tr(H_i H^k)= 
\delta_i^k$.
We then proceed to rewrite $A_{\beta}$ in (\ref{Iko})  as follows:
\begin{eqnarray}
A_{\beta} & = & \frac{1}{2} \sum_{\al,\al+\bt \in \Phi^+}
\tr( [E_{-\al},E_{-\bt}] E^{-\al-\bt} )
\tr( [E_{-\al-\bt}, E_{\bt} ] E^{-\al} ) \nonumber \\ 
 & = & \frac{1}{2} \sum_{\al,\al+\bt,\g \in \Phi^+}
\tr( [E_{-\al},E_{-\bt}] E_{\g} )
\tr( [E^{\g}, E_{\bt} ] E^{-\al} ) \nonumber  \\
 & = & \frac{1}{4} \sum_{\al \in \Phi^+}
\tr( [E_{-\al},E_{-\bt}] [E_{\bt},E^{-\al} ] ) \nonumber  \\
 &  & + \frac{1}{4} \sum_{\al,\g \in \Phi^+}
\tr( [E_{-\bt},E_{\g}] E_{-\al} )
\tr( E^{-\al} [  E^{\g} , E_{\bt}] )\nonumber \\
 & = & \frac{1}{4} \sum_{\al \in \Phi^+}
\tr( [E^{-\al},[E_{-\al},E_{-\bt}]] E_{\bt} )\nonumber  \\
 &  & + \frac{1}{4} \sum_{\g \in \Phi^+, \g \neq \bt}
\tr( [E^{\g},[E_{\g}, E_{-\bt}]] E_{\bt} ) \nonumber \\
 & = & \frac{1}{4} c_2({\cal G}) \tr(E_{-\bt} E_{\bt})\nonumber \\
 &  & - \frac{1}{4} \sum_{i}
\tr( [H^i,[H_i, E_{-\bt}]] E_{\bt} ) \nonumber \\
 &  & - \frac{1}{4} 
\tr( [E_{\bt},E_{-\bt}] [E_{\bt},E^{\bt} ] )\nonumber  \\
 & = &  {c_2(\G) \over 2 \vert \beta \vert^2\vert}
 - {1\over 2 \vert \beta\vert^2} \sum_{i} \beta(H_i) \beta(H^i)
-   {\vert \beta \vert^2 \over 8} \tr( H_\beta H_\beta )
\nonumber \\
&=& {c_2(\G) \over 2 \vert \beta \vert^2\vert} -1.
%\label{jan1}
\nonumber
\end{eqnarray}
In the last step we used that $\vert \beta \vert^2 = 
\sum_i \beta(H_i)\beta(H^i)$, which holds simply 
by the definition of the scalar product of the roots, and 
that $\tr(H_\beta H_\beta) = 4/\vert \beta \vert^2 $, which
follows \cite{H} from  (\ref{normal}).

\end{document}